\documentclass[prd,twocolumn,aps,amsmath,nofootinbib,superscriptaddress]
{revtex4}

\usepackage{graphicx}
\usepackage{bm}

\newcommand{\nn}{\nonumber} 
\newcommand{\bea}{\begin{eqnarray}}
\newcommand{\eea}{\end{eqnarray}}
\newcommand{\comment}[1]{}
\newcommand{\Mp}{m_{\rm P}}
\newcommand{\hMp}{\hat{m}_{\rm P}}
\newcommand{\hmu}{\hat{\mu}}
\newcommand{\mS}{{\mathcal S}}
\newcommand{\mI}{{\mathcal I}}
\newcommand{\mV}{{\mathcal V}}
\newcommand{\mA}{{\mathcal A}}

\newcommand{\mO}{{\mathcal O}}

\begin{document}


\preprint{CALT 68-2619} 
\preprint{RUNHETC-2006-31}

\title{The scale of gravity and the cosmological constant within a  
landscape}

\author{Michael L.~Graesser}
\affiliation{California Institute of Technology, Pasadena, CA 91125}
\affiliation{Department of Physics, Rutgers University, Piscataway, 
NJ 08540}
\author{Michael P.~Salem}
\affiliation{California Institute of Technology, Pasadena, CA 91125}


\begin{abstract}
It is possible that the scale of gravity, parameterized by the apparent 
Planck mass, may obtain different values within different universes in an 
encompassing multiverse.  We investigate the range over which the Planck 
mass may scan while still satisfying anthropic constraints.  The window 
for anthropically allowed values of the Planck mass may have important 
consequences for landscape predictions.  For example, if the likelihood 
to observe some value of the Planck mass is weighted by the inflationary 
expansion factors of the universes that contain that value, then it 
appears extremely unlikely to observe the value of the Planck mass that
is measured within our universe.  This is another example of the runaway 
inflation problem discussed in recent literature.  We also show that the 
window for the Planck mass significantly weakens the anthropic constraint 
on the cosmological constant when both are allowed to vary over a 
landscape.
\end{abstract}


\maketitle

\section{Introduction}
\label{sec:introduction}

Theoretical results from inflationary cosmology~\cite{eternal} and from 
string theory~\cite{strings,bdg} motivate the possibility of an eternally 
inflating multiverse that is populated by an infinite number of 
sub-universes, each obtained via local tunneling, diffusion, and/or 
classical slow-roll into one of a myriad of allowed meta-stable states.  
(For other motivations to consider such a landscape see for example 
refs.~\cite{otherlandscapes}.)  In this landscape picture, each of these 
universes may contain different values for physical parameters, or even 
different particles and interactions, than those that are observed 
within our local universe.  However, the anthropic principle~\cite{anth} 
asserts that the physical laws that may be observed within any universe 
must be restricted to those that permit the evolution of observers in the 
first place.  

Combined with the anthropic principle, the landscape picture has emerged 
as a plausible explanation for many striking features of our universe.
In particular, it has been used to justify the ``unnatural'' smallness of 
the Higgs mass~\cite{anthhiggs}, the cosmological 
constant~\cite{anthcc,anthcc2}, and the neutrino masses~\cite{anthnu}; to 
predict the size of supersymmetry breaking~\cite{bdg,dinethomas}; to 
describe the tilt in the spectrum of density perturbations~\cite{anthtilt} 
and other inflationary parameters~\cite{Tegmark:2004qd}; to constrain the 
baryon to photon ratio~\cite{Linde:1985gh} and the ratio of baryons to dark 
matter~\cite{Linde:1987bx,TARW}, as well as to explain some seemingly 
fine-tuned relationships between parameters describing the theories of 
quantum electrodynamics and quantum chromodynamics~\cite{anthQEDQCD}.  
Nevertheless, generating precise predictions from a landscape picture 
faces several major challenges.    

One of these challenges is to identify an anthropic criteria that is 
both specific and compelling.  Yet even after such an anthropic 
condition has been defined, it is a daunting task to discern its 
environmental requirements, to deduce their implications for physical 
parameters, and then to derive the associated anthropic constraints.  
One may proceed by considering the variation in only one physical 
parameter, starting from its value within our universe.  However, 
apparently tight constraints on any single parameter may be 
significantly weakened when more than one parameter is allowed to vary.  
This seems to be the case with both the Higgs mass~\cite{weakless} (see 
however ref.~\cite{weaklesscounter}) and the cosmological 
constant~\cite{varyingQ1,varyingQ2}.  Moreover, all anthropically 
allowed universes may not be connected by the continuous variation of 
physical parameters.  For example, the seemingly viable ``cold big bang'' 
universe~\cite{Aguirre:2001zx} results from independently varying several 
cosmological parameters to values very far from those obtained within 
our universe.

In addition, to calculate the expectation values of physical parameters 
within a landscape requires determining an appropriate measure to 
weight among the various possible 
universes~\cite{probs1,probs2,probs3,Linde:2006nw}.  
That is, a precise anthropic criteria does not account for all of the 
selection effects that contribute to the probability for a particular 
universe to be observed.  Universes may be more or less likely based on 
how readily they are obtained via the physical dynamics that govern the 
multiverse.  To account for this requires a complete understanding of the 
multiverse, its landscape, and the governing theory.  Indeed, there is a 
more subtle challenge underlying this program, which is to develop an 
appropriate and self-consistent calculus to regulate calculations 
involving the infinite number of infinitely expansive universes that may
be contained within the multiverse~\cite{probs1,probs2,probs3,Linde:2006nw}.

Nevertheless, a set of hypotheses to resolve these challenges may be 
excluded if it predicts a very low likelihood to observe a universe with 
some physical characteristic that our universe possesses.  For example, 
consider a proposal that includes a specific notion of observer, a 
consistent calculus to determine expectation values over the multiverse, 
and a theory to describe the landscape including how meta-stable states 
are mapped onto universes within the multiverse.  If this proposal then 
predicts that an exponentially small number of observers measure a 
cosmological constant at or below the value obtained in our universe, 
then the proposal and the specific landscape in question are probably not 
both correct.  

We investigate the possibility that the scale of gravity may scan over 
the landscape. This is consistent with the results of 
ref.~\cite{Arkani-Hamed:2005yv}, where only parameters with mass 
dimension were found to vary over a model of the landscape.  We 
everywhere parameterize the scale of gravity using the (reduced) Planck 
mass $\Mp$.  Although $\Mp$ is commonly taken to be a fixed fundamental 
scale, this need not be the case.  For example, the multiverse may be 
governed by a low-energy theory with the Lagrangian,
\bea\phantom{\Bigg(\Bigg)}
{\mathcal L} = \frac{1}{2} \sqrt{-g} M^2 F[\phi] R + 
{\mathcal L}_\phi+{\mathcal L}' \,, 
\label{lel}
\eea  
where the fundamental mass scale is $M$, $R$ is the Ricci scalar, $\phi$ 
is the collection of fields that specify the meta-stable state of a 
universe, ${\mathcal L}_\phi$ is the effective Lagrangian for these fields, 
and ${\mathcal L}'$ is the effective Lagrangian for matter.  We assume that 
within each meta-stable state the fields $\phi$ are very massive and fixed 
to values $\phi\to\phi_*$ and are therefore non-dynamical.  The strength of 
gravity will therefore be a constant within each meta-stable state, 
determined by the effective Planck mass,
\bea\phantom{\Bigg(\Bigg)}
\Mp \equiv \sqrt{F[\phi_*]}M \,. 
\eea
We also assume that within each universe $\Mp$ is fixed prior to slow-roll 
inflation.

Alternatively, our analysis may be viewed in the so-called Einstein 
frame where the scale of gravity is everywhere fixed.  To accomplish 
this, one simply performs the conformal transformation 
$g_{\mu\nu}\to F^{-1}[\phi_*]\tilde{g}_{\mu\nu}$.  Then the scale of 
gravity is everywhere $M$, but all other parameters with dimension mass 
are scaled by the factor $F[\phi_*]^{-1/2}$.  Thus our analysis is 
equivalent to fixing the Planck mass to be the fundamental scale of 
physics but varying all other mass scales uniformly.  Stated another 
way, in our analysis the frame-independent ratio of masses $m/\Mp$ 
scales as $F[\phi_*]^{-1/2}$, where for example $m$ may be the cutoff 
of the theory, the Higgs mass, or the scale of strong dynamics.  
This picture was 
previously suggested in the penultimate section of ref.~\cite{weakless}.
The idea that the effective value of $\Mp$ may vary across the multiverse 
within the context of Brans-Dicke theory  was studied in 
refs.~\cite{Linde:1989tz,probs2}.  In addition, ref.~\cite{Biswas:2005vz} 
studied a model of the form of Eq.~(\ref{lel}) to show how inflationary 
dynamics can explain the hierarchy between the apparent Planck scale and 
the electroweak scale.  As described below, our focus is different from 
the focus of this work.    

We calculate the range over which $\Mp$ may scan while still satisfying 
anthropic constraints.  For completeness we consider a wide range of
environmental constraints.  These relate to halo, galaxy, and star 
formation, in addition to galactic and stellar dynamics.  
We restrict attention to universes that possess 
the same particles, interactions, couplings, and physical scales that are 
observed within our universe.  Note that this means the cut-off to the 
low-energy effective theory depends on $M$, not $\Mp$.  In addition, 
we assume that whatever mechanisms drive inflation and provide 
bareogenesis are unchanged (except insofar as they depend on $\Mp$) 
across the landscape that we consider.
Within the context of the model described by Eq.~(\ref{lel}), these 
assumptions require the existence of a large number of states that have
approximately equal particle physics parameters yet different values of 
$\Mp$.  We do not explore the interesting case of a landscape model that
permits only correlated changes in $\Mp$ and the other particle physics
parameters.  Note that the above restrictions are conservative in the 
sense that lifting them can only expand the range of allowed $\Mp$.

Many anthropic constraints relate to the formation of galaxies and 
depend on the spectrum of energy density fluctuations evaluated at 
matter-radiation equality.  On any given distance scale, this spectrum 
has an approximately Gaussian distribution about some root mean square 
(rms) amplitude, and we calculate constraints as if all 
fluctuations have the rms amplitude.  Perhaps not surprisingly, the range 
of $\Mp$ that is consistent with all of the anthropic constraints is 
rather narrow.  It also depends on what models are chosen for inflation, 
baryogenesis, and the dark matter.  As an example, if we assume that 
inflation is chaotic with potential 
$V(\varphi)=\frac{1}{2}m_\varphi^2\varphi^2$, that baryogenesis results 
from efficient leptogenesis, and assume a weakly interacting massive 
particle (WIMP) to be the dark matter, then anthropic considerations 
combine to constrain $\Mp$ to be $0.1\lesssim\hMp\lesssim 1.5$, where 
$\hMp$ is the ratio between $\Mp$ and the value obtained within our 
universe, about $2.4\times 10^{18}$ GeV.  

Even a very narrow anthropic range for $\Mp$ may have significant
consequences for proposals to calculate its expectation value within a
landscape.  In particular, it is plausible that some proposals will
ultimately weight universes in part according to their inflationary
expansion factor.  This expansion factor depends exponentially on the
number of e-folds of inflation that the universe undergoes, which in
turn depends on $\Mp$.  In this case the probability distribution for 
$\Mp$ will be peaked only where some other selection effect cancels this 
strong exponential dependence.  This other selection effect could be a 
very sharp peak or boundary to the underlying landscape distribution; 
otherwise the effect must come from an exponentially strong anthropic 
dependence on $\Mp$.  Yet such a strong anthropic dependence on $\Mp$ 
would be in conflict with the observation that $\Mp$ has even a narrow
anthropic window in our universe.  Thus we are forced to conclude that 
under these weighting schemes the observation of our universe is either 
extremely atypical or our value of $\Mp$ sits at some sharp peak or 
boundary in the underlying landscape distribution.  This point is 
completely analogous to the ``$\sigma$-problem'' and ``$Q$ catastrophe'' 
identified in refs.~\cite{FHW}.

We here note that a runaway problem associated with varying the 
effective Planck mass during eternal inflation has already been discussed
in refs.~\cite{Linde:1989tz,probs2}.  These papers studied the evolution 
of $\Mp$ in Brans-Dicke theory when the Brans-Dicke field is allowed to 
be dynamical during inflation.  On the other hand, we study the case 
where the fields $\phi$ in Eq.~(\ref{lel}) are very massive and therefore 
non-dynamical.  The difference between these scenarios is subtle because 
at some level $\phi$ must be dynamical in order for the landscape to be 
populated within the multiverse.  Our approach is to assume that the 
fields $\phi$ are only dynamical at the very high energies that dominate 
the dynamics of the multiverse.  At these energies the dynamics of $\phi$ 
could be described as in refs.~\cite{Linde:1989tz,probs2} or they could 
be described by different effects.  We simply treat these dynamics as unknown 
except to assume that the $\phi$ are fixed prior to the slow-roll 
inflation that eventually reheats into each of the anthropically 
favorable low-energy universes such as our own.

We also consider the anthropic window for the cosmological constant 
$\Lambda$ when both $\Lambda$ and $\Mp$ are allowed to (independently) 
scan over the landscape.  Even when the allowed range for $\Mp$ is 
relatively narrow, it still allows for a significant broadening of the 
allowed range for $\Lambda$.  To see this, note that $\Lambda$ is 
constrained only by Weinberg's anthropic bound~\cite{anthcc},
\bea
\rho_\Lambda\lesssim \rho_{\rm eq}^{\phantom{3}}\sigma_{\rm eq}^3 \,.
\label{WeinbergLambda}
\eea
Here $\rho_\Lambda$ is the energy density in cosmological constant,
$\rho_{\rm eq}$ is the matter density at matter-radiation equality, and 
$\sigma_{\rm eq}$ is the typical fluctuation in matter density at 
equality.  The broadening occurs because for WIMP dark matter, 
decreasing $\Mp$ significantly increases $\rho_{\rm eq}$ and for most 
models also significantly increases $\sigma_{\rm eq}$.  For example, if 
we again assume chaotic inflation with potential 
$V(\varphi)=\frac{1}{2}m_\varphi^2\varphi^2$ and that baryogenesis 
results from efficient leptogenesis, then $\rho_{\Lambda}$ may be over 
a million times the value observed within our universe when 
$\hMp \gtrsim 0.1$.  Of course, a larger anthropic window for 
$\rho_\Lambda$ does not necessarily imply that our value of 
$\rho_\Lambda$ is less likely to be observed.  We illustrate the 
distribution of observed values of $\rho_\Lambda$ with a very
simplified calculation.  The results of this calculation suggest that to 
observe the cosmological constant at or below the level obtained within 
our universe is very unlikely unless the landscape distribution of $\Mp$ 
is dominated by values very near to or larger than the value obtained 
within our universe.

The remainder of this paper is organized as follows.  In 
Section~\ref{sec:anthropic} we calculate the range of $\Mp$ allowed by 
anthropic constraints in universes otherwise like ours.  Constraints come 
from a variety of cosmological processes and we summarize our results in 
Section~\ref{sec:summary}.  Then in Section~\ref{sec:prior} we argue that 
the value of $\Mp$ that we observe is extremely unlikely if universes 
within the landscape are weighted by their inflationary expansion factor.  
In this section we also discuss some caveats to this argument.  The 
scenario where both $\Mp$ and $\Lambda$ may vary across the landscape is 
discussed in Section~\ref{sec:cc}.  Finally, we draw our conclusions in 
Section~\ref{sec:conclusions}.

\section{Anthropic constraints on the scale of gravity}
\label{sec:anthropic}

It is straightforward to organize the immediate effects of changing the 
scale of gravity when all other mass scales and couplings are kept fixed 
(this implies that the cut-off of the theory is also fixed).  Then 
scanning the Planck mass corresponds to changing the proportionality 
constant between the Einstein and the stress-energy tensors, 
\bea
G_{\mu\nu}= \Mp^{-2}\,T_{\mu\nu} \,.
\eea
In a homogeneous universe this simply changes the relationship between 
the Hubble rate $H$, its time rate of change $\dot{H}$, and the energy 
($\rho$) and pressure ($P$) densities,
\bea
H^2=\frac{\rho}{3\Mp^2} \,,\quad \dot{H} = -\frac{\rho+P}{2\Mp^2} \,.
\eea
This bears upon anthropic conditions because the Hubble rate determines 
when particle interactions freeze out of equilibrium.  This affects the 
relative densities of, for example, matter to radiation and protons to 
neutrons.   

Of course the universe is only approximately homogeneous.  According to 
the present understanding, inhomogeneities are generated by quantum 
fluctuations in at least one scalar field as it exits the Hubble radius 
during (nearly) de Sitter expansion in the early universe.  If this is the 
case, then the Hubble rate also effects the size of the initial 
inhomogeneities.  At late times, these inhomogeneities re-enter the Hubble 
radius and the scale of gravity takes on a new role.  Then gravity 
provides a self-interaction to over-densities that may cause them to grow.  
Over-densities that grow too large become gravitationally bound and 
separate from the cosmic expansion.  Within these structures, the 
expansion of the universe is inconsequential but the scale of gravity 
still determines the internal dynamics.   

We analyze the anthropic significance of these effects in chronological
order, beginning with the effects on inflation.  We then discuss 
baryogenesis, big bang nucleosynthesis, matter domination, structure 
formation, stellar dynamics, and finally the stability of stellar systems.  
The anthropic constraints are displayed in Fig.~\ref{fig1} and summarized 
in Section~\ref{sec:summary}.  The complexity of this analysis, along with 
the many uncertainties in our understanding of various cosmological 
processes, make a precise determination of anthropic constraints 
impractical.  Therefore we strive for approximations that capture the key 
effects of scanning $\Mp$.  Usually, we determine the dominant scaling 
behavior of a quantity with $\Mp$ and cite a precisely determined value 
from our universe to determine the value in another universe.  Unless 
otherwise stated, the values for cosmological parameters within our 
universe are taken from the tables in ref.~\cite{TARW} (note however that
we work in terms of the reduced Planck mass).  Throughout this paper we 
denote the ratio of a quantity to the value that it obtains within our 
universe using a hat, for example
\bea
\hMp \equiv \frac{\Mp}{2.4\times10^{18}{\rm GeV}} \,.
\eea
Finally, we use units where $\hbar=c=k_B=1$.

\subsection{Inflation}
\label{sec:inflation}

An early period of inflation is believed to have homogenized our universe 
and yet provided the seeds of cosmic structure through the generation of 
small density perturbations (for reviews of inflation see for 
example~\cite{inflationreview}).  We parameterize these effects using the 
total number of e-folds of inflation $N$ and the Bardeen curvature 
perturbation $\zeta$.  In principle, both $N$ and $\zeta$ are constrained 
by anthropic considerations.  Meanwhile, for inflation to occur in the 
first place requires that the Hubble radius expand at a rate slower than 
the speed of light.  This effect is parameterized by requiring that the 
first slow-roll parameter, $\epsilon_I\equiv -\dot{H}/H^2$, is smaller 
than unity.  Finally, at some point inflation must end and the universe 
must reheat to establish the initial conditions for the subsequent big 
bang evolution.  We discuss the $\Mp$ dependence of each of these below.

\subsubsection{Satisfying Slow-Roll for $N$ e-folds of Inflation}

For inflation to occur in the first place requires that at some time 
$\epsilon_I< 1$.  When inflation is driven by the potential energy of a 
canonical scalar field $\varphi$, the first slow-roll parameter can be 
written 
\bea
\epsilon_I \simeq \frac{\Mp^2}{2}\left( \frac{V_\varphi}{V}\right)^2 \,,
\eea
where $V$ is the inflaton potential and the subscript on $V_\varphi$ 
denotes differentiation with respect to $\varphi$.  Although at first 
glance $\epsilon_I$ appears to increase with increasing $\Mp$, this can
usually be compensated for by starting the inflaton $\varphi$ further up 
the potential.  This is the case with each of the canonical inflationary 
models presented below.  Therefore we assume that the occurrence of 
inflation in the first place does not significantly constrain $\Mp$.    

Meanwhile, the total duration of inflation is constrained by the need to 
homogenize a universe large enough to allow for the formation of structure.  
This constraint, however, is very weak, since our observable universe 
appears immensely larger than is necessary to form a galaxy.  
Moreover, inflationary scenarios that predict the observed value of 
$\zeta$ typically allow for far more e-folds of inflation than are 
necessary to enclose our universe.  Therefore we assume that the 
anthropic bound on $N$ does not significantly constrain $\Mp$.

\subsubsection{The Curvature Perturbation $\zeta$}

Anthropic constraints on the amplitude of $\zeta$ stem from primordial 
black hole production, structure formation, and the stability of stellar 
systems and are described in Sections~\ref{sec:equality}, 
\ref{sec:structure}, and \ref{sec:closeencounters}.
Presently we discuss the dependence $\zeta(\Mp)$ for future reference.  
For inflation driven by a canonical scalar $\varphi$, the curvature 
perturbation on a co-moving scale with wave-vector $k$ is  
\bea
\zeta(k) \propto \frac{V^{3/2}}{\Mp^3V_\varphi}\bigg|_{k = aH}\,,
\label{zeta}
\eea
where $V$ and $V_\varphi$ are evaluated when the scale $k$ exits the 
Hubble radius. Anthropic constraints on $\zeta$ apply to scales 
$k\lesssim k_{\rm eq}$, where $k_{\rm eq}$ is the wave-vector of the 
Hubble radius at matter-radiation equality.  The potential $V$ 
evaluated when these scales first exit the Hubble radius may depend on 
$\Mp$ even when $V(\varphi)$ does not.  

In our universe, $\zeta$ does not change appreciably with $k$, and we 
assume this holds at least approximately in other universes.  Therefore we 
take $\zeta\approx\zeta(k_{\rm eq})\equiv\zeta_{\rm eq}$ over all scales 
of interest.  The scale $k_{\rm eq}$ of matter-radiation equality itself 
depends on $\Mp$.  However this dependence is logarithmic and its effect 
on $\zeta_{\rm eq}$ is suppressed by the smallness of the slow-roll 
parameter, so we ignore it.  Still, to solve for $\zeta_{\rm eq}$ requires 
to choose a specific model for inflation.  Since there is no standard model 
of inflation, we must be content with a only a plausible range for the 
dependence on $\Mp$.  We deduce this range by studying several of the 
most popular models of inflation.  The results for chaotic 
inflation~\cite{chaotic} with $V(\varphi)\propto\varphi^p$, for hybrid 
inflation~\cite{hybrid}, for natural inflation~\cite{natural}, and for 
ghost inflation~\cite{ghost} are listed in Table~\ref{table1}.  

\begin{table}
\begin{tabular}{ l  l }
\hline\hline\vspace{1pt}
Mechanism to generate $\zeta_{\rm eq}$\quad\quad\quad\, & 
\quad $\Mp$ dependence of $\zeta_{\rm eq}$ \\
\hline
Inflation with $V(\varphi)\propto\varphi^p$ & 
\quad $\Mp^{p/2-2}$ \\
Natural inflation & 
\quad $\Mp^{-3}$ \\
Hybrid inflation & 
\quad $\Mp^{-3}$ \\
Ghost inflation &
\quad $\Mp^{-5/2}$ \\
End of inflation scenario & 
\quad $\chi\,\Mp^{-3}$ \\
Curvaton scenario & 
\quad $\chi^{-1}\Mp^{-1}$ \\
Inhomogeneous reheating & 
\quad $\chi^{-1}\Mp^{-1}$ or $\chi\,\Mp^{-1}$ \\
\hline\hline
\end{tabular}
\caption{The dependence of $\zeta_{\rm eq}$ on $\Mp$ for a variety 
of mechanisms to generate the curvature perturbation.  The result for
inhomogeneous reheating depends on the relative size between $\chi$ 
and other mass scales in the Lagrangian, and can interpolate between 
the two dependences given above.}
\label{table1}
\end{table}

The curvature perturbation $\zeta_{\rm eq}$ may also be generated at the 
very end of inflation~\cite{endofinf} or even much later as in the 
curvaton~\cite{curvaton} and inhomogeneous reheating~\cite{DGZK} 
scenarios.  In each of these models a sub-dominant scalar $\chi$ 
receives fluctuations while the fluctuations to the inflaton are presumed 
to be negligible.  The fluctuations in $\chi$ are then transferred to 
radiation either at the end of inflation, during reheating, or during a 
phase transition much later.  In each of these cases, the amplitude of the 
curvature perturbation depends on the local vacuum expectation value 
(vev) of $\chi$.  The dependence of $\zeta_{\rm eq}$ on $\chi$ and 
$\Mp$ when each of these mechanisms operates efficiently (that is, 
when any reheating occurs far out of equilibrium) is listed in 
Table~\ref{table1}.  

We note that these models are very flexible to anthropic selection.  
Although in principle $\chi$ may be set by interactions such that it is 
fixed among the set of universes we consider, in most cases $\chi$ is a 
stochastic variable over these universes.  When $\zeta_{\rm eq}$ is 
generated at the end of inflation or via the curvaton or inhomogeneous 
reheating scenarios, then this implies that $\zeta_{\rm eq}$ is also a 
stochastic variable over different universes.  Therefore in these cases 
there exist universes with far different $\Mp$ but the same 
$\zeta_{\rm eq}$ as in our universe, as well as universes with the same 
$\Mp$ but different $\zeta_{\rm eq}$.

For future convenience we write $\zeta_{\rm eq}$ in the form,
\bea
\hat{\zeta}_{\rm eq}\approx \hMp^{-\alpha} \,,
\label{zetanote}
\eea
where according to our notation $\hat{\zeta}_{\rm eq}$ is the curvature 
perturbation relative its value in our universe.  The various models 
of inflation that we studied suggest that we should restrict $\alpha$ to 
the range $0\leq\alpha\leq 3$.  However, the mechanisms that generate 
$\zeta_{\rm eq}$ at the end of or well after inflation may generate a 
wide range of $\zeta_{\rm eq}$ for a wide range of $\Mp$.  Although it is 
technically possible that keeping inflationary parameters fixed but 
varying $\Mp$ will cause the dominant contribution to the curvature 
perturbation to shift from one mechanism to another, this scenario should 
still be well approximated by the above guidelines so long as the 
variation in $\Mp$ is not too large.

\subsubsection{Reheating}

The reheating of the universe after inflation is achieved by coupling the 
inflaton to other degrees of freedom.  In a typical model, after inflation 
the inflaton rocks within its potential well and red-shifts like matter.  
The radiative decay products of the inflaton then dominate the energy 
density of the universe only after the Hubble rate falls below the decay 
width $\Gamma_I$.  At this time the energy density in the inflaton is 
$\rho_I=3\Mp^2\Gamma_I^2$.  If $\Gamma_I$ is independent of $\Mp$, then 
the reheat temperature scales like
\bea
\hat{T}_{RH} = \hMp^{1/2} \,.
\eea
We have no empirical knowledge about reheating other than that $T_{RH}$ 
is above the temperature of big bang nucleosynthesis.  However, if our 
universe is described by a grand unified theory, then $T_{RH}$ must be 
below the temperature of monopole production.  In addition, if net baryon 
number is not generated during reheating, then $T_{RH}$ must be high 
enough to support the dominant mechanism of baryogenesis.  We comment on 
constraints like these in Section~\ref{sec:prior}.

\subsection{Baryogenesis}
\label{sec:baryogenesis}

We parameterize the net baryon number of a universe with the ratio 
between the number density of baryons and the number density of photons: 
$\eta\equiv n_b/n_\gamma$.  Although we see no direct anthropic 
constraints on the value of $\eta$, it will enter into the anthropic 
constraints described in Sections~\ref{sec:structure} and 
\ref{sec:closeencounters}.  Presently, we seek to parameterize the 
dependence of $\eta$ on $\Mp$ for future reference.  As with inflation, 
there is no standard model of baryogenesis.  Therefore we must again 
content ourselves with only a range for the dependence on $\Mp$, based 
on the most plausible mechanisms.  For a summary of these see for 
example the reviews of refs.~\cite{baryogenesis}.

Perhaps the most plausible mechanism to produce net baryon number is 
leptogenesis~\cite{leptogenesis}.  For example, net lepton number is 
rather easily obtained by the out-of-equilibrium decay of a right-handed 
neutrino (RHN).  The resulting lepton asymmetry can then be converted 
into net baryon number by sphaleron transitions within the Standard 
Model~\cite{sphaleron}.  The value of $\eta$ that results from 
leptogenesis depends on how far out of equilibrium the RHN decays.  
If RHN decay occurs far out of equilibrium then the resulting baryon 
asymmetry $\eta$ is independent of $\Mp$.  Otherwise, it scales roughly
according to $\eta\propto\Mp^{-1}$.  Note that baryogenesis via 
leptogenesis requires that the RHN, for example, be produced in the first
place.  We comment on this requirement in Section~\ref{sec:prior}.

The Standard Model of particle physics itself generates appropriate 
conditions for baryogenesis, when it is augmented by relatively light 
supersymmetric (SUSY) scalars to strengthen the electroweak phase 
transition.  The process of electroweak baryogenesis is complex; yet 
interestingly it operates independently the scale of gravity.  That is,
although universal expansion is necessary to decrease the temperature of
the universe and thus spur the electroweak phase transition, this process
is relatively independent the rate of temperature change.  Therefore 
electroweak baryogenesis gives a baryon asymmetry that is independent of 
$\Mp$~\cite{baryogenesis}.  

Finally, we look at Affleck Dine (AD) baryogenesis~\cite{Affleck:1984fy}.  
This mechanism takes advantage of scalar fields that possess baryon or 
lepton number, as would exist in SUSY or a grand unified theory (GUT).  
During inflation these fields may acquire large vevs, and then the 
influence of baryon non-conserving interactions on their subsequent 
evolution may generate significant baryon number.  There are many models 
to implement AD baryogenesis and each may give a different dependence on 
$\Mp$.  We simply give the result for a set of scenarios described in 
ref.~\cite{Dine:1995kz}, where the scalar fields overlap a SUSY flat 
direction that is lifted by non-renormalizable interactions and a negative 
induced mass term during inflation.  To fit the resulting baryon asymmetry 
to observation requires different parameterizations when different 
non-renormalizable interactions dominate; however in each case the 
dependence on $\Mp$ is given by 
$\eta\propto\Mp^{-3/2}$~\cite{Dine:1995kz}.

For future reference, it is convenient to write the baryon to photon
ratio in the form
\bea
\hat{\eta}\approx\hMp^{-\beta} \,.
\label{etanote}
\eea
From the above discussion, we expect $0\leq\beta\leq3/2$, with perhaps 
the most plausible values being $\beta=0$ or $3/2$.  As was the case with
inflation, if more than one mechanism contributes to the net baryon number 
then we still expect the $\eta$ that results to be well approximated by 
the above guidelines.

\subsection{Big Bang Nucleosynthesis}
\label{sec:bbn}

The process of big bang nucleosynthesis (BBN) populates the universe with 
light elements.  In particular, the mass fractions of hydrogen $(X)$ and 
helium $(Y)$ are important for anthropic considerations described in 
Sections~\ref{sec:structure} and \ref{sec:stellard}.  Since we are only 
interested in $X$ and $Y$, we may take a very simplified view of BBN.  
Specifically, we assume that BBN generates appreciable concentrations of 
only hydrogen and helium-4.  This is clearly appropriate within our 
universe, where the other products of BBN account for only about 0.01\% 
of the mass fraction of the universe.  Although this fraction may change 
significantly for differing values of $\Mp$, it would take a very large 
variation in $\Mp$ for this change to become significant next to $X$ or 
$Y$.  A basic description of BBN can be found in ref.~\cite{bbcref}.

When we approximate BBN to result in only hydrogen and helium-4, we 
require only the ratio of neutrons ($n$) to protons ($p$) to deduce 
$X$ and $Y$.  Specifically, 
\bea
X\approx \frac{1-n/p}{1+n/p} \,,\qquad Y\approx1-X \,.
\label{Xeq}
\eea
A free neutron has $\Delta E\approx 1.3$ MeV more energy than a free 
proton.  Thus if neutrons and protons are kept in thermal equilibrium by 
interactions that convert each into the other, then $n/p$ is given by 
the Boltzmann factor, 
\bea
n/p = \exp\left( -\Delta E/T \right) \,,
\eea
where $T$ is the temperature.  Neutrons and protons are converted into 
each other via interactions such as $n+\nu\longleftrightarrow p+e$, where 
the $\nu$ denotes an electron neutrino and the $e$ an electron.  In fact, 
the rates of these interactions decrease faster with temperature than 
does the Hubble rate.  Therefore, below some temperature $T_F$ the 
universe expands too rapidly for, for example, an $n$ and a $\nu$ to 
find each other and convert into a $p$ and an $e$.  Below this 
temperature the relative concentrations of $n$ and $p$ are fixed and the 
interaction is said to freeze-out.\footnote{The decay of neutrons 
decreases $n/p$ from its value at $T_F$ by about 14\% within our universe.  
This loss is determined by the time at which neutrons are efficiently 
captured into Helium, which depends on $\Mp$ only via a logarithmic 
dependence on $\eta(\Mp)$~\cite{bbcref}.  Inspecting Fig.~4.4 of 
ref.~\cite{bbcref}, it can be shown that in the most extreme case of 
$\beta=3/2$, less than half of the neutrons decay for $\hMp\lesssim 6$.  
On the other hand, for $\hMp\gtrsim 6$ the helium fraction is less than 
a tenth of the hydrogen fraction.  Therefore this effect is always 
negligible at our level of analysis.}   

The freeze-out temperature $T_F$ is obtained by equating the total 
rate of interactions converting neutrons to protons, $\Gamma_{\rm np}$, 
with the Hubble rate $H$.  Then the neutron to proton ratio is
$n/p\approx\exp\left( -\Delta E/T_F\right)$.  Our universe contains
$n/p\approx 1/7$ such that $X\approx 3/4$.  Note that varying $\Mp$ so as 
to decrease $T_F$ works to decrease $n/p$ and therefore push $X$ closer 
to unity.  Since in our universe $X\approx 3/4$, this effect is negligible 
at our level of analysis.  On the other hand, for temperatures 
$T\geq T_F$ the rate of conversion between neutrons and protons is 
$\Gamma_{\rm np}\propto T^5$ and is independent of $\eta$.  Since 
$H\propto T^2/\Mp$, the freeze-out temperature scales like 
$T_F\propto\Mp^{-1/3}$ when $T_F$ is larger than in our universe.  
Therefore we find, 
\bea
n/p\approx\exp\left[-\ln (7)\,\hMp^{1/3}\right] \,.
\eea  
The fractions $X$ and $Y$ are given by Eqs.~(\ref{Xeq}).  For example, 
when $\hMp=10$, 5, 0.2, and 0.1 we have $X\approx 1$, 0.9, 0.5, and 0.4, 
respectively.

\subsection{Matter Domination}
\label{sec:equality}

We have assumed that the early universe is dominated by relativistic 
degrees of freedom, at least since BBN.  However as the universe cools, 
massive degrees of freedom eventually become non-relativistic.  As it
becomes non-relativistic, the number density of this matter becomes 
exponentially suppressed relative that of radiation.  However, this 
dilution eventually causes matter to freeze out of equilibrium with 
the remaining radiation.  Subsequently, the energy density of a massive 
species $i$ will redshift as $\rho_i\propto m_in_i$ while the energy 
density in radiation scales as $\rho_{\rm rad}\propto Tn_{\rm rad}$.  
Thus it is inevitable that matter should ultimately come to dominate the 
energy density of the universe (structure formation constraints ensure 
that the cosmological constant does not become significant before matter 
domination).

Nevertheless, the energy density at matter-radiation equality and the 
fraction of matter in baryons are relevant to anthropic constraints 
described in Sections~\ref{sec:structure} and \ref{sec:closeencounters}.
In addition, our assumption that the early universe is radiation 
dominated does not hold if $\zeta$ is too large.  In this case, 
primordial black holes may dominate the energy density of the universe 
while baryons are still relativistic.  Then all of the baryons would be 
redshifted away or swallowed into black holes.  This possibility is 
studied at the end of this section.  In the following we neglect the 
neutrino content of the universe.  Their influence on cosmology is 
commonly viewed as insignificant and we do not expect this to change 
since as a hot relic their density relative baryons is fixed. 
In addition, we assume the dark matter to be a WIMP.  This allows for 
relatively precise predictions, as opposed to, for example, axion dark 
matter where the density is set by a stochastic variable~\cite{axion}.  
(Note however that the stochastic nature of axion dark matter makes this 
possibility more flexible to anthropic selection, see for example 
refs.~\cite{Linde:1987bx,TARW}.)

The energy density in a WIMP dark matter candidate is set by the relic 
abundance that results from the freeze-out of annihilation interactions 
when the temperature drops below the mass of the WIMP.  After 
matter-radiation equality this gives the scaling, 
\bea
\rho_{\rm cdm}\propto \Mp^{-1}\,T^3 \,.
\eea
Meanwhile, after baryogenesis the relative abundance of baryons $\eta$ 
is conserved.  Therefore at temperatures below the nucleon mass the 
energy density in baryons scales like 
\bea
\rho_b\propto \eta T^3 \,.
\eea  
Using that in our universe $\rho_b/\rho_{\rm cdm}\approx 1/5$ and
that the energy density in radiation scales as $\rho_{\rm rad}\propto T^4$,
we find the energy density at matter-radiation equality to be
\bea
\hat{\rho}_{\rm eq}\approx
\left( \frac{1}{6}\hat{\eta}+\frac{5}{6}\hMp^{-1} \right)^{\!4} \,.
\eea
Finally, we note the baryon fraction within matter, 
\bea
\hat{f}_b\approx 
\left( \frac{1}{6}+\frac{5}{6}\hat{\eta}^{-1}\hMp^{-1} \right)^{\!-1} \,. 
\label{baryonf}
\eea   

As described above, these results do not hold if $\zeta$ is so large as 
to produce an abundance of primordial black holes (PBHs)~\cite{pbh}.  
Numerical analysis reveals that a PBH is formed when an energy density 
fluctuation $\sigma\gtrsim 0.7$ enters the Hubble 
radius~\cite{Niemeyer:1999ak}. Meanwhile, during radiation domination 
$\sigma=\frac{4}{3}\,\zeta$ at Hubble radius 
crossing~\cite{Mukhanov:1990me}.  Therefore we require 
$\zeta(k)\lesssim 0.5$ in order to prevent the formation of a PBH when 
the scale $k$ enters the Hubble radius.  In fact, this implies a somewhat 
stronger constraint on $\zeta$.  This is because $\zeta$ is a 
stochastic variable with a Gaussian tail and because a PBH need not be 
formed each Hubble time in order for PBHs to dominate the energy density 
of the universe.  This constraint is worked out in ref.~\cite{varyingQ1} 
and we follow that analysis.    

The likelihood that a curvature perturbation with root mean square 
$\zeta$ is greater than or equal to $1/2$ is 
$f(\zeta) = {\rm erfc}(2^{-3/2}\zeta^{-1})$, where the 
complementary error function is defined 
${\rm erfc}(x)\equiv 2\pi^{-1/2}\int_x^\infty e^{-z^2}dz$.
PBHs redshift like matter while the other degrees of freedom redshift 
like radiation.  Therefore by matter-radiation equality PBHs will 
compose roughly $(a_{\rm eq}/a_{\rm pbh})f$ of the energy density of 
the universe.  Here $a_{\rm pbh}$ is the scale factor at which newly 
formed PBHs have sufficient mass to persist until equality.  Then 
PBHs do not dominate the energy density of the universe when 
\bea
{\rm erfc}(2^{-3/2}\zeta_{\rm eq}^{-1})\lesssim 
\frac{a_{\rm pbh}}{a_{\rm eq}}\,,
\label{nopbh1}
\eea
where we have neglected any tilt in $\zeta$.  This approximation 
underestimates the largest $\zeta_{\rm eq}$ since recent observations 
suggest a negative tilt~\cite{wmap}.

The erfc function depends very strongly on its argument; therefore the 
$\Mp$ dependence of the ratio $a_{\rm pbh}/a_{\rm eq}$ is inconsequential 
for our analysis.  As one looks to earlier times, the lifetimes of PBHs
decreases more rapidly than cosmic time decreases.  Therefore 
Eq.~(\ref{nopbh1}) ensures that PBHs dominate at no time prior to 
equality.  Solving for when PBH lifetimes equal about 70,000 years gives
$a_{\rm pbh}/a_{\rm eq}\sim 10^{-20}$ which gives 
$\zeta_{\rm eq}\lesssim 6\times10^{-2}$.  Translating this into a 
constraint on $\Mp$, we find
\bea
\hMp^{\alpha}\gtrsim 7\times 10^{-4}\,,
\label{nopbh}
\eea  
where we have used that the density fluctuation at Hubble radius crossing
is $\sigma\approx 5\times 10^{-5}$~\cite{wmap}.  Eq.~(\ref{nopbh}) is 
always weaker than the stellar lifetime constraint of 
Section~\ref{sec:stellard}.

\subsection{Structure Formation}
\label{sec:structure}

The formation of structure within our universe occurs in several stages.  
First, over-densities in the nearly pressureless dark matter begin to grow 
upon entry into the Hubble radius.  During radiation domination, this growth 
is logarithmic with time, while after the dark matter comes to dominate the 
energy density of the universe over-densities grow in proportion to the 
growth in the cosmic scale factor.  On the other hand, over-densities in
the baryons cannot grow until after recombination. However, within an 
e-fold or so after recombination they have grown to match the over-density 
in dark matter, and subsequently grow in proportion to the growth in the 
cosmic scale factor.  When these over-densities have grown sufficiently 
they separate from the Hubble flow and virialize to form what are termed 
halos.   

After virialization, the cold dark matter within halos is stabilized 
against gravitational collapse by its inability to release its kinetic 
energy.  However, the baryons within the halo must collapse beyond their 
initial virialization radius if they are to fragment and condense into 
galaxies and ultimately into stars. This requires that the baryons have 
a means to dissipate their thermal energy.  The constraints on $\Mp$ 
that are implied by these stages of structure formation are discussed in 
the sections below.  Presently, we describe the initial growth in 
over-densities for future reference.  The subject of galaxy formation 
and in particular star formation is complex and not yet fully understood.  
We rely heavily on the simplifying assumptions and models of 
refs.~\cite{varyingQ1,TARW}.    

We find it convenient to track the evolution of over-densities in 
position space, as opposed to Fourier space.  At matter-radiation 
equality, the variance of energy density fluctuations over scales with 
co-moving radius $R$ is,    
\bea
\sigma_{\rm eq}^2(R) = \int_0^\infty\!\! 
\frac{dk}{2\pi^2}\,k^2\,W(kR)T(k)P(k) \,,
\label{sigmavariance}
\eea
where $W$ is a window function that may be chosen to be a ``top hat'' 
with radius $R$, $T$ is a transfer function to account for the evolution
of perturbations between when they enter the Hubble radius and equality, 
and $P$ is the primordial power spectrum of fluctuations, 
$P(k)\propto\langle\zeta(k)^2\rangle$.  We parameterize a co-moving 
scale with radius $R$ according to the total mass $\mu$ that is enclosed 
within a sphere of radius $R$.  In addition, we measure $\mu$ relative to 
the mass of our galaxy (more precisely the mass of our galaxy plus its 
dark matter halo); thus $\mu=1$ corresponds to $10^{12}M_\odot$, where 
$M_\odot$ is the mass of the sun.  

A numerical curve fit to Eq.~(\ref{sigmavariance}) gives~\cite{TARW}, 
\bea
\sigma_{\rm eq} \simeq 1.45\times 10^{-3}\,
s(\mu)\hat{\zeta}_{\rm eq}(\Mp) \,,
\label{delta}
\eea
where the function $s(\mu)$ carries the scale dependence of 
$\sigma_{\rm eq}$.  This scale dependence occurs because at the time of 
equality smaller scales have been within the Hubble radius for a longer 
time than larger scales.  The function $s(\mu)$ is equivalent to 
Eq.~(A13) in ref.~\cite{TARW}.  However, we have normalized $s(\mu)$ such 
that $s(1)=1$.  In addition, we define the variable $\mu$ with respect to 
a different scale than the authors of ref.~\cite{TARW}.  Therefore within 
this paper $s$ is given by,
\bea
s(\mu)\!&=&\!\Big[ 
\left(0.76\ln[17+\mu^{-1/3}]-0.22\right)^{-0.27} \Big. \nn\\
& & \quad +\,\, \Big. 0.17\mu^{0.18}\,\, \Big]^{-3.7} \,.
\label{sdef}
\eea
Note that $s$ is a decreasing function of $\mu$.  In addition, we 
emphasize that $\sigma_{\rm eq}$ is the root mean square (rms) value of a 
Gaussian random field.  Therefore constraints involving $\sigma_{\rm eq}$ 
(or $\hat{\zeta}_{\rm eq}$) are never sharp in the sense that they may be 
overcome by fluctuations that happen to be larger or smaller than is 
typical. 

After recombination but before the domination of cosmological constant, 
a linear over-density is given by~\cite{anthnu}
\bea
\sigma\approx
\left(\frac{2}{5}+\frac{3}{5}\frac{a}{a_{\rm eq}} \right)
\sigma_{\rm eq} \,.
\label{growth}
\eea
Soon after recombination the first term is negligible.  
Eq.~(\ref{growth}) is accurate until cosmological constant domination, 
after which $\sigma$ grows by another factor of about 1.44 and then 
stops.  An over-density separates from the Hubble flow and virializes 
when a linear analysis gives $\sigma= 1.69$~\cite{virialize}.  
Thus the cosmic mean energy density at virialization is,
\bea
\rho_*\approx\left(\frac{3}{5}\frac{\sigma_{\rm eq}}{1.69}\right)^3\!
\rho_{\rm eq}\approx 1.4\times 10^{-10}\rho_{\rm eq}
\hat{\zeta}_{\rm eq}^{\,\,3}s^3 \,.
\eea
The energy density within the condensed halo is larger by roughly a 
factor of $18\pi^2$.  We denote this as,
\bea
\rho_{\rm vir}\approx 18\pi^2\rho_*\approx 
2.4\times 10^{-8}\rho_{\rm eq}\hat{\zeta}_{\rm eq}^{\,\,3}s^3 \,.
\eea   
Note that these quantities depend on both $\Mp$ and the mass scale 
$\mu$ of the virialized halo.  

The above description of halo formation relies on three important 
aspects of the standard cosmology:  we assume that the dark matter 
density dominates over the baryon density, that recombination occurs
before the virialization of the dark matter halo, and that 
virialization occurs before the domination of cosmological constant.  
Enforcing the above conditions implies constraints on $\Mp$.  These
are discussed in the next section.  We discuss the possibility for a 
non-standard path toward structure formation in 
Appendix~\ref{sec:nonstandard}.

\subsubsection{Halo Virialization}
\label{sec:ccconstr}

Before proceeding to galaxy formation, we must ensure that over-densities
separate from the cosmic expansion and virialize before the domination of 
the cosmological constant halts their growth.  As mentioned above, an 
over-density $\sigma$ has separated from the Hubble flow when a linear 
analysis gives $\sigma \geq 1.69$.  On the other hand, the maximum size 
that is reached by an rms linear over-density is given by,
\bea
\sigma_\infty \approx 1.44\times\frac{3}{5}
\frac{a_\Lambda}{a_{\rm eq}}\sigma_{\rm eq}
\approx 3.20\,\hat{\rho}_{\rm eq}^{1/3}\hat{\zeta}_{\rm eq}s \,,
\label{sigmainf1}
\eea  
where $a_\Lambda$ is the scale factor at $\rho_\Lambda$ domination.  
Therefore the requirement that $\sigma_\infty \geq 1.69$ gives 
\bea
\hat{\rho}_{\rm eq}\hat{\zeta}_{\rm eq}^{\,\,3}s^3 \gtrsim 0.1 \,.
\label{lambdaconstraint1}
\eea
Since with a larger value for $\Mp$ structures form later, this is a 
constraint against increasing $\Mp$.  Substituting previous results into 
Eq.~(\ref{lambdaconstraint1}) gives
\bea
\left(\frac{1}{6}\hMp^{-\beta}+\frac{5}{6}\hMp^{-1}\right)^{\!4}\!
\hMp^{-3\alpha} s(\mu)^3 \gtrsim 0.1 \,.
\label{LambdaConstraint}
\eea 
The curves that saturate this inequality for various choices of $\alpha$ 
and $\beta$ are displayed in Fig.~\ref{fig1} under the label 
``$\sigma_\infty=1.69$.''  In light of these plots, our value of 
$\Mp$ may be construed as nearly saturating this constraint.  However, 
this perception derives from the strong $\Mp$ dependence of 
$\rho_{\rm vir}$, and not from our galaxy being at the edge of saturating
the Weinberg bound.  In addition this constraint, analogous to 
many others below, holds for an rms fluctuation $\sigma_{\rm eq}$ but is 
weaker for larger fluctuations.  For these reasons we emphasize that 
although Eq.~(\ref{LambdaConstraint}) does not allow for $\Mp$ to be 
increased significantly, it is still true that our value of $\Mp$ is not 
at the edge of the anthropic range.  Since the existence of an 
anthropically allowed window surrounding our value of $\Mp$ is essential 
to the arguments of Section~\ref{sec:prior}, we provide a more elaborate 
discussion of this boundary in Appendix~\ref{sec:MpLambda}.  This may 
serve as an illustration of how `soft' are other constraints that depend 
on $\sigma_{\rm eq}$. 

The above analysis assumes that halos virialize at least an e-fold or so
after recombination.  This is to ensure that baryons may collapse into 
the dark matter potential wells and participate in the virialization.  
Thus we require $\rho_*/\rho_{\rm rec}\lesssim e^{-3}$.  The energy 
density at recombination is set by the temperature of recombination, 
which depends only logarithmically on $\Mp$ and $\eta$.  We ignore this 
logarithmic dependence and take $T_{\rm rec}\approx 3000$ K in every 
universe that we consider.  Using that at any time after equality 
$\rho_*\propto\rho_{\rm rec}\propto T^3$, we find for this constraint,
\bea
\hat{\rho}_{\rm eq}^{\,-3/4}\,\hat{\zeta}_{\rm eq}^{\,-3} s^{-3}
\gtrsim 8\times 10^{-8} \,.
\eea
Since with a smaller value of $\Mp$ the matter energy density at 
equality and the amplitude of density perturbations are both larger, this 
is a constraint against decreasing $\Mp$.  In terms of $\hMp$ and $\mu$, 
this gives
\bea
\left(\frac{1}{6}\hMp^{-\beta}+\frac{5}{6}\hMp^{-1}\right)^{\!-3}\!
\hMp^{3\alpha} s(\mu)^{-3} \gtrsim 8\times 10^{-8} \,.
\label{virconstr}
\eea
The curves that saturate this constraint for various $\alpha$ and 
$\beta$ are displayed in Fig.~\ref{fig1} with the label
``$t_{\rm rec}=e^{-2/3}t_{\rm vir}$.''  

Finally, we require that dark matter dominate over baryonic matter so 
that dark matter potential wells are deep enough to condense baryon 
over-densities after recombination.  This simply translates into the 
constraint $f_b\lesssim 1/2$, which gives
\bea
\hMp \!&\lesssim&\! 5^{1/(1-\beta)} \quad {\rm for}\,\, \beta < 1 \nn\\
\hMp \!&\gtrsim&\!  5^{1/(1-\beta)} \quad {\rm for}\,\, \beta > 1 \,.
\label{baryonfraction}
\eea
For $\beta=1$ this argument provides no constraint on $\Mp$ since 
in that case $f_b$ is independent $\Mp$.

\subsubsection{Galaxy Formation}
\label{sec:cooling}

Although the dark matter within a halo cannot dissipate its kinetic 
energy to further collapse, the baryons may do so via electromagnetic 
interactions.  If the cooling timescale $\tau_{\rm cool}$ is less than 
the timescale of gravitational dynamics $\tau_{\rm grav}$, then not only
do the baryons collapse, but perturbations in the baryon density fragment 
into smaller structures.  These structures ultimately fall into a 
rotationally supported disk.  Perturbations may further fragment if the 
disk satisfies the Jeans instability criteria, which is that 
$\tau_{\rm grav}$ be less than the time it takes for a pressure wave to 
traverse the perturbation.  Fragmentation continues until perturbations 
become Jeans-stable and over-densities relax adiabatically into hot 
balls of gas.  This appears to be the path by which halos within our 
universe ultimately condense into galaxies of stars (for background see 
for example ref.~\cite{BT}).    

In order to ensure galactic dynamics similar to those within our 
universe, one might therefore first impose that for typical halos, 
$\tau_{\rm cool}\lesssim\tau_{\rm grav}$.\footnote{
We note that more careful considerations involving galactic dynamics may 
suggest a far weaker constraint than the one we pursue.  We require 
$\tau_{\rm cool}\lesssim\tau_{\rm grav}$, where each timescale is 
evaluated at virialization.  However, the baryons within a halo will cool 
even if this condition is not satisfied.  As described in 
ref.~\cite{coolingo}, this cooling pushes the gas of baryons along a 
curve in the temperature-density phase space that eventually leads to 
the condition $\tau_{\rm cool}\lesssim\tau_{\rm grav}$ being satisfied, 
albeit at a much later time.  It is then necessary to consider any
other factors that might constraint the timescale $\tau_{\rm cool}$.  
Ref.~\cite{TARW} has pointed out that if $\tau_{\rm cool}$ is not much
smaller than the Hubble time, then baryons do not cool significantly 
before being reheated by halo mergers.  Eventually halo merging ceases
due to cosmological constant domination.  However, even then, one must
worry about too large a faction of baryons evaporating from the halo
before they cool sufficiently to sink deeper into the gravitational
potential well~\cite{TARW}.  These considerations are beyond the scope
of this paper.}
We take the dynamical timescale of the halo to be the time it would 
take for a test particle to free-fall to the center of the halo.  For a 
spherical halo of constant density, this is 
\bea
\tau_{\rm grav}\approx \sqrt{\textstyle\frac{3}{2}\pi^2}\,
\Mp\rho_{\rm vir}^{-1/2}\,.
\eea
For the cooling timescale we use the total thermal energy divided by the 
rate of energy loss, per unit volume:
\bea
\tau_{\rm cool}\approx\frac{3}{2}\frac{f_b\,\rho_{\rm vir}}{m_N\mu_b}
\frac{T_{\rm vir}}{\Lambda_c} \,,
\eea  
where $\mu_b$ is the mean molecular weight of the baryons in the 
halo (in units of the nucleon mass $m_N$), $T_{\rm vir}$ is the mean 
temperature of the halo, and $\Lambda_c$ is the rate of energy loss per 
unit volume.  The quantity $f_b\,\rho_{\rm vir}/m_N\mu_b$ is the baryon 
number density, including electrons.    

The mean molecular weight depends on the ionization fraction and the 
hydrogen mass fraction of the halo.  For example, for a fully ionized 
halo we have,
\bea
\mu_b\approx\frac{n_{\rm H}+4n_{\rm He}}{n_e+n_{\rm H}+n_{\rm He}}
\approx \frac{4}{3+5X} \,,
\eea
where the subscripts denote electrons, hydrogen, or helium.  Note that 
$\mu_b$ never strays more than a factor of two from unity.  To estimate
the temperature of the halo, we first note that in a virialized halo the
mean kinetic energy equals half the mean gravitational binding energy.  
Thus for a halo of mass $M$ we write 
$Mv_{\rm vir}^2\approx\frac{3}{40\pi}\Mp^{-2}M^2R^{-1}$, 
where $v_{\rm vir}$ is a characteristic velocity for virialized particles 
and $R$ is the radius of the halo.  Since 
$M\approx\frac{4\pi}{3}R^3\rho_{\rm vir}$ and since 
$T_{\rm vir}\approx\frac{1}{3}\mu_bm_Nv_{\rm vir}^2$, we obtain
\bea
T_{\rm vir}\approx \frac{1}{78}\mu_b m_N\Mp^{-2}M^{2/3}\rho_{\rm vir}^{1/3} \,.
\label{virialT}
\eea  
Note that $T_{\rm vir}\propto\sigma_{\rm eq}$ so that $T_{\rm vir}$ is
a stochastic variable for halos of a given mass.

The baryons within a halo may cool via Compton scattering, bremsstrahlung,
the excitation of hydrogen or helium lines, in addition to other 
mechanisms.  These all contribute to the rate of thermal energy 
dissipation $\Lambda_c$.  Thus $\Lambda_c$ is a complicated function of 
temperature, which also depends on the halo composition and therefore 
the hydrogen and helium fractions $X$ and $Y$.  Rather than attempt an 
estimate of $\Lambda_c$, we use the cooling rates given in 
refs.~\cite{cooling}.  These include the processes listed above, but we 
neglect the possibility for molecular cooling, which is insignificant at 
the temperatures we consider.  The galactic cooling constraint 
$\tau_{\rm cool}\lesssim\tau_{\rm grav}$ is now,
\bea
\Mp^3\,\Lambda_c\,f_b^{-1}M^{-2/3}\rho_{\rm vir}^{-11/6} 
\gtrsim 5\times 10^{-3} \,.
\label{coolingconstraint}
\eea
The curves that saturate this inequality are displayed in Fig.~\ref{fig1} 
with the label ``$\tau_{\rm cool}=\tau_{\rm grav}$.''  

As noted in ref.~\cite{TARW}, if a galaxy contains too little mass then
early supernovae may blow away a significant fraction of its baryons when
they explode.  We expect the effects of a supernova to be relatively 
localized if the gravitational binding energy of the galaxy by far 
exceeds the energy released in the supernova~\cite{TARW}.  This can be
ensured by requiring that the energy released in the supernova be less
than the halo binding energy, 
\bea
E_{\rm bind}\approx\frac{3}{40\pi\Mp^2}\frac{M^2}{R}
\approx \frac{1}{26}\Mp^{-2}M^{5/3}\rho_{\rm vir}^{1/3} \,.
\label{halobind}
\eea
Note that the baryons within a galaxy are much more tightly bound than
when in the original halo (see for example the galactic disk estimates 
of Sections~\ref{sec:starformation} and~\ref{sec:closeencounters}).  
Therefore Eq.~(\ref{halobind}) is a significant underestimate of the 
binding energy of a galaxy.

We expect the energy released in a supernova to scale roughly as the 
binding energy of a Chandrasekhar mass at its Schwarzschild 
radius~\cite{TARW}, or as the binding energy of a typical star, both of 
which are proportional to $\Mp^3$ (see Section~\ref{sec:stellard}).  Thus 
we write this energy $E_{\rm sn}\hMp^3$, where $E_{\rm sn}\approx 10^{51}$ 
erg is the typical supernova energy within our universe.  Requiring that
$E_{\rm bind}\gtrsim E_{\rm sn}$ gives,
\bea
\hMp^{-5}\hat{\rho}_{\rm eq}^{1/3}\hat{\zeta}_{\rm eq}\,\mu^{5/3} s
\gtrsim 4\times 10^{-9} \,.
\label{presn}
\eea 
Since halos of a given mass become more weakly gravitationally bound as 
$\Mp$ is increased, Eq.~(\ref{presn}) is a constraint against increasing 
$\Mp$.  After inserting previous results this becomes,
\bea
\left(\frac{1}{6}\hMp^{-\beta}+\frac{5}{6}\hMp^{-1}\right)^{\!\!4/3}\!
\hMp^{-5-\alpha}\mu^{5/3} s(\mu) \gtrsim 4\times 10^{-9} \,.\,\,\,
\label{supernova}
\eea
The curves that saturate Eq.~(\ref{supernova}) are displayed in 
Fig.~\ref{fig1} under the label ``$E_{\rm bind}=E_{\rm sn}$.''

\subsubsection{Star Formation}
\label{sec:starformation}

If the above conditions are met, the baryons in a halo will radiate away
energy and settle into a disk supported by its angular momentum.  We 
then require that the disk fragment so that ultimately stars may 
form~\cite{TARW}.  The stability of galactic disks against both radial 
and vertical perturbations can be studied using a standard Jeans 
analysis, which compares the dynamical timescale $\tau_{\rm grav}$ to the
time it takes a pressure wave to traverse the perturbation.  It turns out 
that the stability criteria for the two modes differ by only an order 
unity coefficient~\cite{BT,Fall:1980sj,TARW}.  In ref.~\cite{Fall:1980sj}
it is shown that for perturbations in the vertical direction, 
perturbations are unstable when the total mass of the 
disk satisfies, 
\bea
M_{\rm disk}\gtrsim 120\,\Mp^2 v_pv_c R_{\rm disk} \,,
\label{diskconstraint1}
\eea
where $v_p$ is the typical peculiar velocity of particles in the disk, 
$v_c$ is the circular velocity of these particles, and $R_{\rm disk}$
is the disk radius.  

The mass of the disk is simply the mass of the baryons in the halo, 
$M_{\rm disk}=f_b M$.  Meanwhile, the peculiar velocity is related to 
the temperature of the disk.  This temperature will be the lowest 
temperature to which the baryons can cool as they collapse, which is 
roughly set by the hydrogen line temperature $T_{\rm H}\approx 10^4$ 
K.\footnote{Although the rate of galactic cooling is reduced below 
the temperature of hydrogen line freeze-out, cooling still proceeds via 
molecular transitions in, for example, $H_2$.  Therefore $T_{\rm min}$ 
may become very small if one is willing to wait a long time before disk 
fragmentation.  We follow ref.~\cite{TARW} and study when galactic 
dynamics are similar to those within our universe.}
Therefore the typical peculiar velocity may be written, 
\bea
v_p \approx \sqrt{\frac{3T_{\rm H}}{\mu_bm_N}} \,.
\label{peculiarv}
\eea
The circular velocity $v_c$ is deduced by conserving angular momentum as 
the baryonic halo collapses.  On the one hand, the disk angular momentum
can be roughly written as $R_{\rm disk}M_{\rm disk}v_c$.  On the other 
hand, the baryons in a halo start with angular momentum 
$\frac{1}{\sqrt{8\pi}}f_b\lambda\Mp^{-1}M^{3/2}R^{1/2}$, where $\lambda$ 
is the dimensionless spin parameter~\cite{spinparameter},
\bea
\lambda\equiv \sqrt{8\pi}\frac{JE_{\rm bind}}{\Mp M^{5/2}}\,,
\eea
where $J$ is the magnitude of the angular momentum, $E_{\rm bind}$ is 
the gravitational binding energy, and all of the above quantities are 
evaluated for the original halo. Before the gravitational collapse of 
the baryons out of the dark matter halo, it is reasonable to assume 
that the angular momentum of the baryons and dark matter are equally 
distributed according to mass,  such that initially $J_b = M_b J_h /M_h$ 
where the subscripts $b$ and $h$ refer to baryonic and dark matter halo 
quantities \cite{Fall:1980sj}. Then assuming that angular momentum is 
conserved as the baryons in the halo collapse, we find
\bea
v_c \sim \frac{\lambda}{\sqrt{8\pi}}\,
\Mp^{-1}M^{1/2}R^{1/2}R_{\rm disk}^{-1} \,.
\label{circularv}
\eea
We do not require to solve for $R_{\rm disk}$ after this 
expression for $v_c$ is inserted into Eq.~(\ref{diskconstraint1}).  

Eq.~(\ref{diskconstraint1}) can now be written in the simple 
form~\cite{TARW},
\bea
f_b \gtrsim 4\lambda\left(\frac{T_{\rm H}}{T_{\rm vir}}\right)^{\!1/2} \,.
\label{diskinst1}
\eea
The spin parameter $\lambda$ is different for different halos, but it 
is roughly independent of $\Mp$, $\mu$ and the amplitude 
of density perturbations, and it typically lies near 
$\lambda\approx 0.05$~\cite{spinparameter}.  Substituting into 
Eq.~(\ref{diskinst1}) gives
\bea
\hat{\mu}_b^{1/2}\hat{f}_b\hMp^{-1}\hat{\rho}_{\rm eq}^{1/6}
\hat{\zeta}_{\rm eq}^{\,1/2}\mu^{1/3}s^{1/2} \gtrsim 0.2 \,.
\label{discinstability}
\eea
This results in a constraint against increasing $\Mp$.  We use previous 
results to convert this into a constraint on $\hMp$ and $\mu$, which
gives 
\bea
\left(\frac{1}{6}\hMp^{-\beta}+\frac{5}{6}\hMp^{-1}\right)^{\!\!-1/3}\!\!
\hMp^{-1-\beta-\alpha/2}\mu^{1/3}s^{1/2} \gtrsim 0.2 \,,
\eea
where $\hat{\mu}_b(\hMp)$ depends relatively weakly on $\Mp$ and has 
been ignored.  The curves that saturate this constraint for various 
$\alpha$ and $\beta$ are displayed in Fig.~\ref{fig1} with the label 
``disk inst..''      

It is essential that at some point fragmentation ceases so that 
over-densities can smoothly collapse into a star.  The process of 
fragmentation may be seen to terminate when individual fragments become 
sufficiently opaque so as to trap most of their radiation~\cite{starf}.  
In the Jeans picture, this allows the temperature of a perturbation to 
rise and correspondingly increase the sound speed and thus prevent 
further fragmentation.  As described in ref.~\cite{starf}, the mass 
scale at which this occurs is relatively independent the dominant 
contributions to the cooling rate and opacity and gives a typical stellar 
mass that scales like $\Mp^3$.  Interestingly, this is the same scaling 
behavior that restricts the sizes of stars based on their internal 
temperature being high enough to fuse hydrogen and their radiation 
pressure being low enough so as to not blow the star apart.  We elaborate
on this in the next section.

\subsection{Stellar Dynamics}
\label{sec:stellard}

We have so far ensured that the fragmentation of over-densities persists 
on all scales greater than a relatively small scale that is roughly 
proportional to $\Mp^3$.  We now require that the temperature within 
some of the remaining structures is sufficient to fuse hydrogen to form 
a star.  It is possible that the mere existence of stars is not a 
sufficient condition for the existence of observers.  Therefore we also
consider the requirement that some of these stars supernova in order to
generate heavy elements.  In addition, we consider the requirement that
some of these stars have both a surface temperature within a factor of 
two that of the sun and a main-sequence lifetime of at least a few 
billion years.  Our motivation for selecting these specific criteria is
simple.  Without knowing what are the necessary conditions for observers 
to arise within a stellar system, we study what seem at least to be 
two sufficient conditions.

Ref.~\cite{anthQEDQCD} has studied the basic requirements that constrain 
the properties of stars.  For decreasing stellar mass, the central 
temperature must be above some minimum temperature $T_{\rm nuc}$ that is 
necessary to fuse hydrogen.  The central temperature within a low-mass 
star is estimated by balancing the influences of gravitational pressure 
and electron degeneracy.  This gives for the central temperature of a 
low-mass star~\cite{anthQEDQCD},
\bea
T_c \propto \Mp^{-4}M^{4/3} \,,
\eea
where $M$ is the stellar mass.  The least massive stars have 
$T_c=T_{\rm nuc}$.  Since $T_{\rm nuc}$ is independent of $\Mp$, in any 
universe these stars have mass
\bea
M_{\rm min}\propto \Mp^3 \,.
\label{Mmin}
\eea  

On the other hand, there is also an upper limit to the mass of a star.  
If the radiation pressure within a star well exceeds the gravitational 
pressure that its mass can provide, then the star itself becomes 
unstable upon the ignition of its core.  This constrains the maximum 
mass that a star may have to satisfy the scaling~\cite{anthQEDQCD}, 
\bea
M_{\rm max} \propto \Mp^3 \,.
\label{Mmax}
\eea
Note that the minimum and maximum mass of a star both scale as $\Mp^3$.  
This is also the scaling of the typical mass that becomes sufficiently 
opaque to prevent further fragmentation.  This means that, given a window 
of masses for which stars exist in our universe, there will also be such
a window within universes with significantly different values of $\Mp$.

\subsubsection{Stellar Lifetimes and Spectra}

We now seek constraints to ensure that some of the stars produced within 
a particular universe have surface temperatures and main-sequence 
lifetimes that appear to be sufficient for the evolution of observers.  
Our purpose in investigating this condition is to ensure that we do not
overlook what might be viewed as an important anthropic constraint.  
Therefore we adopt a very restrictive perspective and require that some
stars have surface temperatures of at least $3500$ K and that these stars
have main-sequence lifetimes greater than the timescale of biological 
evolution, $\tau_{\rm evol} \approx 5\times 10^9$ yrs.  This surface 
temperature is chosen in part to simplify the calculation of stellar 
lifetimes and in part because a black body at this temperature radiates 
a significant fraction of its power into the frequency band accessible to 
chemistry.  The evolutionary timescale $\tau_{\rm evol}$ should be 
understood to include the time required for a planet to condense and 
cool, minus the time it takes the star to reach main-sequence hydrogen
burning.  This time may be different for different planets, but we do 
not expect it to form the dominant contribution to $\tau_{\rm evol}$.  

The main-sequence lifetime of a star is roughly equal to the available 
energy of the star divided by the typical rate that it radiates energy 
away,
\bea
\tau_\star\propto XML^{-1} \,.
\label{starlifetime}
\eea
Here $L$ is the typical luminosity of the star during main-sequence and 
$X$ is the hydrogen fraction.  We assume that differences in composition 
other than differences in the hydrogen fraction have benign consequences.  
In addition, we neglect the $\Mp$ dependence of the mean molecular weight 
$\mu_\star$, since $\mu_\star$ changes by only roughly a factor of two as 
$X$ ranges from zero to unity.  For an introduction to the major concepts 
of stellar astrophysics, we have found useful refs.~\cite{Limber,CG,KW}.

Our narrow purpose allows for a simplified analysis of the necessary 
stellar dynamics.  Since we specify stars by their surface temperature, 
we write the luminosity $L\propto R^2 T_s^4$ for stellar radius $R$ and 
surface temperature $T_s$.  To eliminate $R$, we note that the central 
temperature of a star scales as 
\bea
T_c\propto \frac{1}{\Mp^2}\frac{M}{R} \,. 
\label{centraltemp}
\eea
Thus we can write,
\bea
\tau_\star\propto \frac{X\Mp^4T_c^2}{MT_s^4} \,.
\label{starform}
\eea  
The lifetime of a star is maximized by considering the minimum allowed
surface temperature, in this case $T_s=3500$ K.  Since both the lower end
and the upper end of the window for stellar masses scale as $\Mp^3$, as a 
basic approximation we may take this to be the scaling for all stellar 
masses at fixed $T_s$ and $T_c$.  Combining this with 
Eqs.~(\ref{starlifetime}), (\ref{centraltemp}), and (\ref{starform})
gives,
\bea
\hat{\tau}_\star \approx \hat{X}\hMp \,,
\eea
where $\hat{\tau}_\star$ is measured in units of the main-sequence 
lifetime of these stars within our universe.  This lifetime is roughly
100 billion years~\cite{CB}.  Therefore the constraint 
$\tau_\star\gtrsim\tau_{\rm evol}$ becomes
\bea
\hat{X}\hMp \gtrsim 5\times 10^{-2} \,.
\label{stellarlife}
\eea  
This requires that $\hMp$ satisfy $\hMp \gtrsim 0.1$.

We now consider this analysis in a little more detail.  In particular, 
we consider the effects of convection and electron degeneracy explicitly 
in order to motivate that we can keep $T_s/T_c$ fixed while scaling $\Mp$ 
and that the stellar mass scales like $\Mp^3$ at fixed $T_c$.  An ionized 
star in which radiation pressure can be neglected and in which the 
energy transport is everywhere dominated by convection is well 
approximated as a polytrope with polytropic index 
$i=3/2$~\cite{CG,Limber}.  This implies certain scalings between stellar 
properties and in particular that for these stars $T_s/T_c$ is independent 
of $\Mp$, $M$ and $R$.  Within our universe, stars with surface 
temperatures at and below $3500$ K have masses $M\lesssim 0.35M_\odot$ 
and are well approximated by these polytropes.  In addition, it can be 
shown that stars defined by these temperatures remain convection 
dominated as $\Mp$ is decreased~\cite{Limber}.  To see this intuitively, 
note that convection is driven by tidal forces and, all else being equal, 
one expects the tidal forces within a star to increase as $\Mp$ is 
decreased.  Therefore, we expect $T_s/T_c$ to be fixed for stars with 
$T_s=3500$ K as $\Mp$ is decreased.

Meanwhile, so long as electron degeneracy is significant within the 
center of the star, the mass required to achieve a fixed central 
temperature scales like $\Mp^3$~\cite{anthQEDQCD}.  Stars with a
surface temperature of $3500$ K are indeed partially degenerate within 
our universe, but they could become non-degenerate after some amount of 
scaling $M \propto \Mp^3$.  To see that there can exist stars for which 
the degeneracy remains fixed, consider again the polytrope model. 
The electron degeneracy at the center of a star is a function of the 
ratio $n_e T_c^{-3/2}$, where $n_e$ is the number density 
of electrons.  The scaling relations applicable to an $i=3/2$ polytrope
imply that the electron density $n_e$ scales like the average density 
of the star, which at constant central temperature scales like 
$\Mp^6M^{-2}$. Next note that the electron degeneracy at the center of a 
star is a constant if the mass of the star scales as $M\propto\Mp^3$ for
fixed central temperature $T_c$.  This is precisely the scaling that 
describes a partially degenerate star, which means that a partially 
degenerate star remains partially degenerate as $\Mp$ is scaled while
keeping $T_c$ fixed.  Therefore we expect stars with a surface 
temperature of $3500$ K to remain partially degenerate as we decrease 
$\Mp$ keeping $T_s$ fixed, so that indeed $M\propto\Mp^3$.  These 
arguments justify the constraint Eq.~(\ref{stellarlife}).

It is illuminating to consider a different form of analysis.  This 
applies to ionized stars where radiation pressure can be neglected but 
radiation dominates over convection in the transport of energy.  Then 
the scaling of the stellar mass $M$ with $\Mp$ for fixed $T_c$ may be 
obtained for a class of stars (so-called ``homologous 
stars''\footnote{This is a very restrictive class, since by definition 
the mass distribution for two homologous stars of mass $M_i$ and 
radius $R_i$ must satisfy $m_1(r)/M_1=m_2(r R_2/R_1)/M_2$, where 
$m_i(r)$ is the mass contained within a sphere of radius $r$.}) 
by applying homologous transformations to the equations of hydrodynamical 
equilibrium.  To perform such an analysis, we must phenomenologically 
model the opacity of the star and the rate of energy generation per
unit mass with the respective formulae~\cite{CG}, 
\bea
\kappa\approx\kappa_0\rho^n T^{-s}\,,\quad
\epsilon\approx\epsilon_0X^2\rho T^{\nu} \,.
\label{opacity}
\eea
The terms $\kappa_0$ and $\epsilon_0$ are constants while the exponents
$n$, $s$, and $\nu$ depend on the physical properties of the star.  
The Kramers opacity of intermediate mass stars such as our sun is modeled 
using $n=1$ and $s=7/2$.  Then it can be shown that, independent of 
$\nu$, the stellar mass scales like $M\propto X^{1/3}\Mp^{10/3}$ for 
fixed central temperature~\cite{CG}. For such stars $T_s/T_c$ is not 
constant, so one cannot use Eq.~(\ref{starform}).  However, at fixed 
central temperature the homology transformations determine the scaling of 
the luminosity to be $L\propto X^2\Mp^6\,M^{-1}\propto X^{5/3}\Mp^{8/3}$. 
Inserting these scalings into Eq.~(\ref{starlifetime}) gives 
\bea
\hat{\tau}_\star'\approx \hat{X}^{-1/3}\Mp^{2/3} \,,
\eea  
where $\hat{\tau}_\star'$ is the main-sequence lifetime in units of
an appropriate lifetime evaluated within our universe. 

This gives a slightly weaker dependence on $\Mp$ than 
Eq.~(\ref{stellarlife}); however stars that are well-described by these 
approximations have shorter lifetimes than those described by an $i=3/2$ 
polytrope.  In addition, the physical characteristics that make these 
approximations applicable will not continue to describe stars as we 
decrease $\Mp$ with $T_s$ fixed.  (They do continue to describe stars as 
$\Mp$ is increased for fixed $T_s$, since this tends to lessen the 
importance of convection.)  Nevertheless, this confirms the qualitative 
form of Eq.~(\ref{stellarlife}) and suggests that an analogous analysis 
would apply if surface temperatures closer to that of the sun were 
demanded.

\subsubsection{Heavy Element Production}

Supernovae are believed to be the exclusive source of heavy elements 
within our universe. However, the dynamics of supernovae are very 
complex and are still not fully understood (for reviews see for example 
refs.~\cite{supernova}).  Therefore ensuring the existence of supernovae
in universes with differing values of $\Mp$ is clearly speculative.  In 
this section we simply provide some qualitative remarks in support of 
this possibility.  There are many types of supernovae within our universe; 
for convenience we focus on what are called type Ia supernovae.  

Type Ia supernovae are understood to erupt via the accretion of matter 
by a white dwarf star.  Meanwhile, white dwarfs are created when a star 
has consumed all of its hydrogen and helium fuel but does not possess 
sufficient mass either to drive its central temperature high enough to 
ignite carbon fusion or to form a black hole.  According to the scaling
relationships discussed in the previous section, the first condition is
always satisfied given that it is satisfied within our universe.  On the 
other hand, the Schwarzschild radius and the physical radius of a star
at fixed central temperature both scale as $R\propto\Mp^{-2} M$.  
Therefore white dwarfs will exist in all of the universes that we 
consider.

The supernova of an accreting  white dwarf proceeds when its growing
mass reaches the Chandrasekhar limit and the star becomes unstable 
through the nuclear ignition of its carbon.  The relevant physical 
scales for this phenomena are set by the Chandrasekhar mass and the 
binding energy of a Chandrasekhar mass at about a Schwarzschild radius.  
Since these and the typical stellar mass all scale as $\Mp^3$, it seems 
plausible that type Ia supernovae would occur within universes with 
significantly differing values of $\Mp$.  This ensures the production of 
heavy elements within these universes.  

There is a possible caveat to this result.  Within the context of another 
anthropic analysis, it has been remarked~\cite{weaklesscounter} that the 
relatively low production of oxygen by type Ia supernovae may 
significantly hinder the formation of life if oxygen is not generated 
elsewhere.  However, it is not clear that this suppression, roughly 
3--8\% relative type II supernova~\cite{sn2}, is sufficient to render life 
overwhelmingly unlikely.  The arguments of ref.~\cite{weaklesscounter} 
were aimed against a scenario where type II supernova would definitely 
not occur.  Since these supernova occur for a wide range of stellar masses 
within our universe, it is plausible that universes with $\Mp$ not too 
unlike ours will also contain type II supernovae.  It is beyond the scope 
of this paper to investigate more precisely for what values of $\Mp$ 
these supernovae will occur.

\subsection{Stability of Stellar Systems}
\label{sec:closeencounters}

As is illustrated in refs.~\cite{varyingQ1,TARW}, an important anthropic 
constraint derives from requiring that stellar systems are stable against
cosmic disruptions.  Specifically, if a second star grazes too close to 
an existing stellar system, then a habitable planet may be thrown out of 
its anthropically fortuitous orbit.  Here we seek a constraint to ensure 
that such encounters are typically too infrequent to interfere with the 
evolution of life.  First, we define a destructive encounter rate, 
\bea
\gamma\sim n_\star\sigma_\star v_p \,,
\eea 
where $n_\star$ is the number density of stars, $\sigma_\star=\pi b^2$ is 
the cross section for an encounter with ``fatal'' impact parameter $b$, and 
$v_p$ is the typical peculiar velocity of a star.  Note that all of the
stars within a given neighborhood have the same circular velocity; thus 
the circular velocity does not contribute to the encounter rate.  

The typical peculiar velocity of the stars in a galaxy is approximately 
determined by the temperature of the constitutive baryons during the 
phase of star formation.  Since the baryons in a galaxy quickly cool 
to about $T_{\rm H}\approx 10^4$ K and cool relatively slowly thereafter, 
we take this to be the relevant temperature.  The corresponding peculiar
velocity is then given by Eq.~(\ref{peculiarv}).  For $\hMp=1$ this gives 
a typical peculiar velocity of about $v_p\simeq 20$ km/s which agrees 
well with observation.  The number density of stars $n_\star$ is equal 
to the number density of baryons divided by the typical number of baryons 
within a star, $N_\star$.  Recall from Section~\ref{sec:structure} that 
the number density of baryons within a galactic halo is equal to 
$f_b\rho_{\rm vir}/m_N\mu_b$.  Therefore $n_\star$ can be written,
\bea
n_\star = \frac{f_\star f_b\,\rho_{\rm vir}}{m_N\mu_b N_\star} \,,
\eea
where $f_\star$ is a fudge factor inserted to account for the increased 
density of the galactic disk, for the fraction of baryons that do not end 
up within stars, and for any clustering that may be involved in the star
formation process.  A typical star within our universe contains $10^{57}$ 
baryons; therefore the results of Section~\ref{sec:stellard} suggest 
$N_\star\sim 10^{57}\hMp^3/\hmu_\star\sim 10^{57}\hMp^3/\hmu_b$.  
To estimate $f_\star$ is somewhat more challenging.

The factor $f_\star$ accounts for several effects.  For example, stars 
may form in clusters such that most stars exist in a neighborhood of 
higher density than the average density of stars in a galaxy.  On the
other hand, a significant fraction of baryons may compose a relatively
diffuse interstellar gas and therefore not contribute to the stellar 
encounter rate.  As it is beyond the scope of this paper to compute the
$\Mp$ dependence of these effects, we simply treat them as being
independent of $\Mp$.  Meanwhile, we also expect $f_\star$ to be 
proportional to the relative density of the galactic disk to that of 
the baryons in the halo.  Rather than concern ourselves with the 
specific geometry of the galactic disk, we study a simple model to 
obtain the $\Mp$ dependence.  We expect the factor $f_\star$ to 
roughly scale like
\bea
f_\star\propto\frac{R^3}{H_{\rm disk}R_{\rm disk}^2} \,,
\eea  
where $H_{\rm disk}$ is the typical disk thickness.  This can be 
written~\cite{Fall:1980sj},
\bea
H_{\rm disk}\propto\Mp^2 M_{\rm disk}^{-1}R_{\rm disk}^2v_p^2 \,.
\eea  
Meanwhile, to solve for $R_{\rm disk}$ we note that the circular 
velocity for stars is given by both by Eq.~(\ref{circularv}) and by,
\bea
v_c \propto\Mp^{-1}M_{\rm disk}^{1/2}R_{\rm disk}^{-1/2} \,.
\eea
Equating these expressions gives $R_{\rm disk}\propto\lambda^2 f_b^{-1}R$.  
Finally, putting all of this together gives,
\bea
f_\star\propto\frac{f_b^5}{\lambda^8}
\frac{v_{\rm vir}^2}{v_p^2} \,.
\label{fstar}
\eea  

Note the strong dependence of $f_\star$ on the spin parameter 
$\lambda$.  The spin parameter is a stochastic variable with statistical 
properties related to those of $\zeta_{\rm eq}$.  For example, the 
Milky Way appears to be characterized by 
$\lambda\approx 0.06$~\cite{MWprop} while typical galaxies may have
$\lambda$ a factor of two larger or smaller than 
this~\cite{spinparameter}.  This and other factors suggest that the 
factor $f_\star$ may vary widely among galactic environments within any 
particular universe.  In addition, as explained above we have ignored 
several effects that might enter into $f_\star$.  For concreteness we 
normalize $f_\star$ to the value that describes our solar environment 
in the Milky Way, $f_\star\sim 10^5$~\cite{TARW}.  This gives
\bea
f_\star \sim 10^5\hMp^{-2}\hat{\mu}_b\hat{f}_b^5\mu^{2/3}
\hat{\rho}_{\rm eq}^{1/3}\hat{\zeta}_{\rm eq}s(\mu) \,,
\eea       
where the dependence of $f_\star$ on $\Mp$ stems entirely from its 
dependence on $f_b$ and $T_{\rm vir}$.  

It is left to calculate the impact parameter for fatal encounters.  
We are specifically interested in the persistence of stellar systems that  
contain a planet in orbit about a star such as those considered in 
Section~\ref{sec:stellard}.  In addition, we focus on planetary orbits 
that receive electromagnetic radiation with an intensity that is 
comparable to that from the sun at the orbit of the earth.  We then 
assume that an encounter will not be devastating to such a stellar 
system if the gravitational field from the grazing star is less than 
a tenth that of the primary star in the vicinity of the orbiting planet.
Therefore we approximate $b$ to be roughly $\sqrt{10}$ times the radius 
of orbit for a planet receiving about the same stellar intensity as the 
earth but in orbit about the stars studied in Section~\ref{sec:stellard}.  
Note that the luminosity of a star is $L\propto R^2 T_s^4$ while the 
intensity at a distance $r$ is $I\propto L/r^2$.  Therefore 
\bea
b=\sqrt{10}\,r_{\rm au}\frac{R_\otimes T_\otimes^2}{R_\odot T_\odot^2}
\frac{R_\ominus}{R_\otimes} \,,
\label{eqb}
\eea    
where the subscript $\odot$ designates a quantity for the sun, 
$\otimes$ designates a star with surface temperature $T_\otimes=3500$ K 
in our universe, $R_\ominus$ is the radius of a star with this surface 
temperature within a universe with a different value for $\Mp$, and 
$r_{\rm au}$ is one astronomical unit.  Eq.~(\ref{eqb}) has been written
so that every quantity can be evaluated within our universe except for 
$R_\ominus/R_\otimes\approx\hMp$, which is deduced using the results 
of Section~\ref{sec:stellard}. 

\begin{figure*}[t!]
\begin{tabular}{c c}
\includegraphics[width=8.5cm]{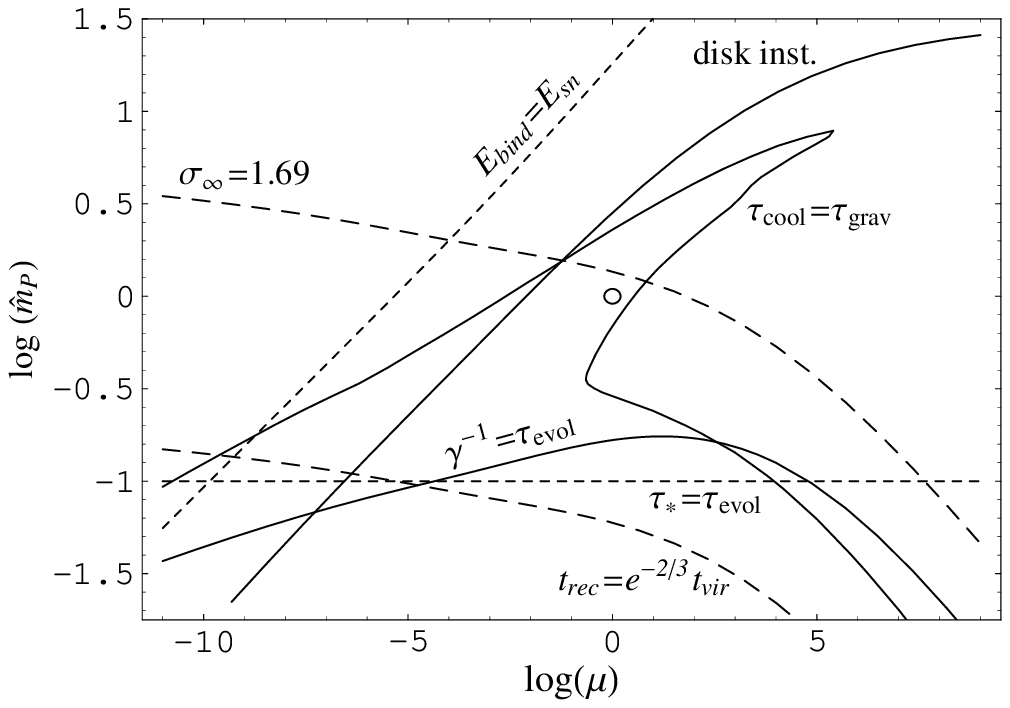} & \quad\quad
\includegraphics[width=8.5cm]{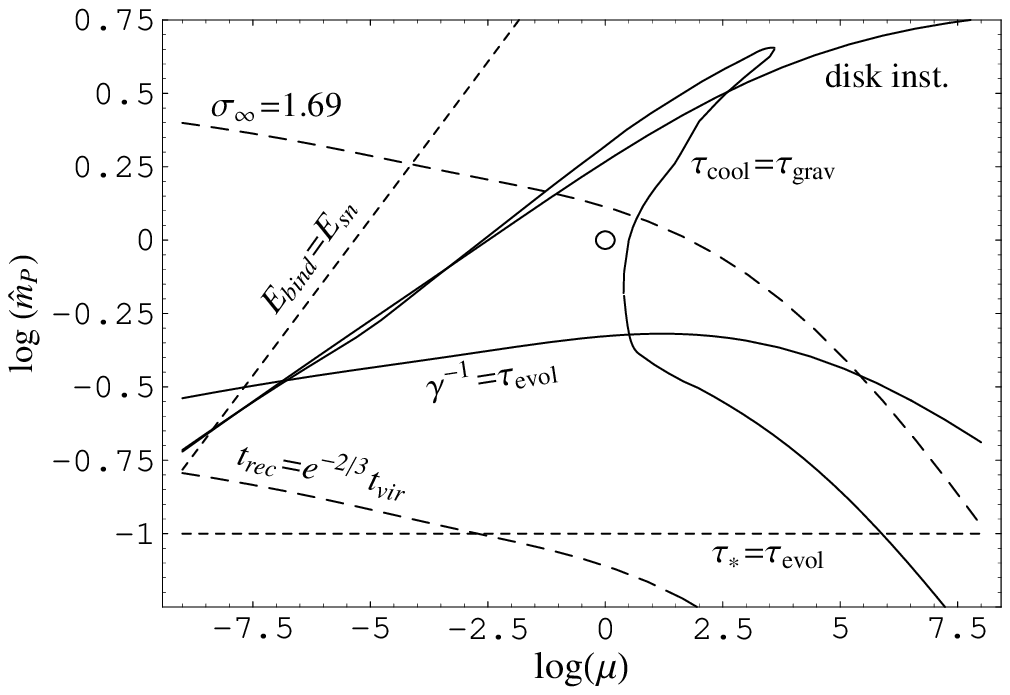} \\
\includegraphics[width=8.5cm]{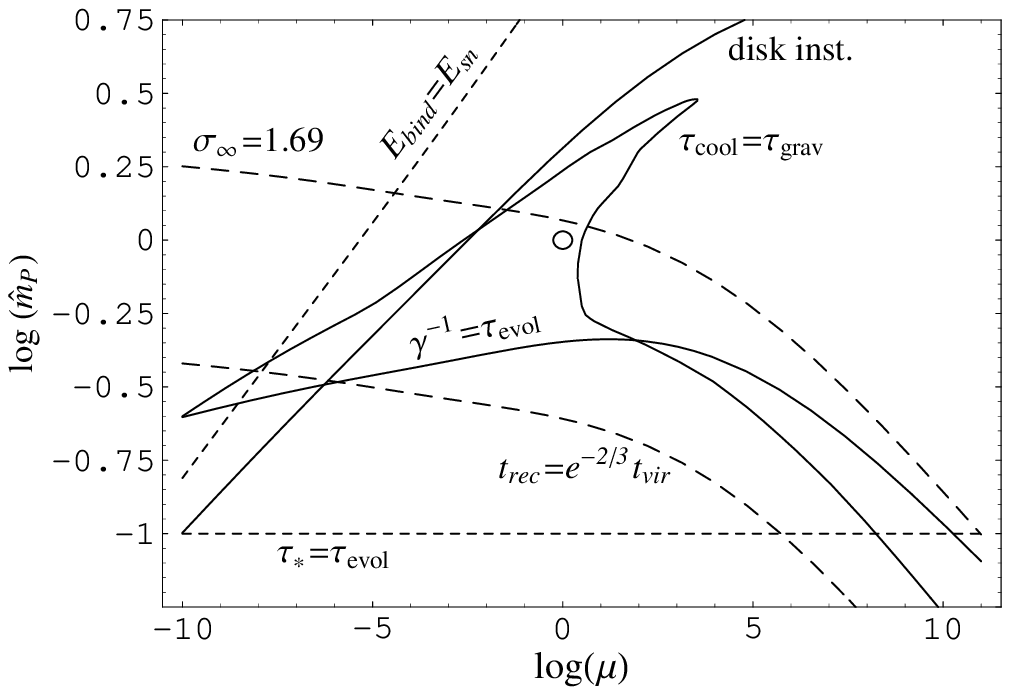} & \quad\quad
\includegraphics[width=8.5cm]{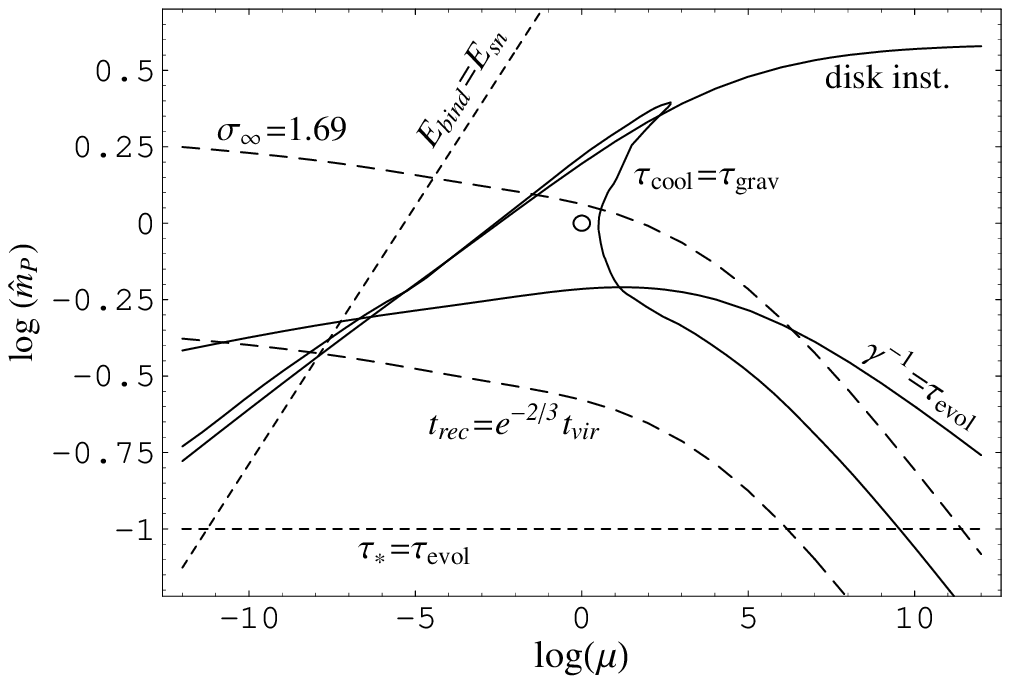} \\
\end{tabular}
\caption{Anthropic constraints on $\hMp$, plot as a function of halo 
mass scale $\mu$, for 
$\alpha=1$, $\beta=0$ (top left panel); 
$\alpha=1$, $\beta=3/2$ (top right panel);
$\alpha=3$, $\beta=0$ (bottom left panel);
$\alpha=3$, $\beta=3/2$ (bottom right panel).  The logarithms are base 
ten and the empty circle corresponds to the mass scale of our galaxy with 
$\Mp$ as observed within our universe.  The region within the $\mu-\hMp$ 
plane that is excluded by any constraint is the region that does not 
include this circle.  The parameters $\alpha$ and $\beta$ as well as the 
labels on the curves are defined in the text (see for example the summary 
of Section~\ref{sec:summary}).  Note that the galactic cooling constraint
does not include the effects of molecular cooling.}
\label{fig1}
\end{figure*}

We put all these results together to obtain $\gamma$.  The constraint
that stellar systems typically survive a dangerous close encounter for 
long enough that life may evolve is $\tau_{\rm evol}\lesssim\gamma^{-1}$,
where again $\tau_{\rm evol}=5\times 10^9$ yrs.  In terms of cosmological 
parameters this is,
\bea
\hMp^3 \hat{\mu}_b^{-1/2}\hat{f}_b^{-6}\hat{\rho}_{\rm eq}^{-4/3}
\hat{\zeta}_{\rm eq}^{\,-4}\mu^{-2/3}s^{-4}\gtrsim 10^{-5}\,,
\label{closeencounters}
\eea 
where we have used the models of ref.~\cite{CB} to substitute
$R_\otimes T_\otimes^2/R_\odot T_\odot^2
\approx\sqrt{L_\otimes/L_\odot}\approx0.14$. Decreasing $\Mp$ reduces 
the cross section for dangerous impacts, since the 
`anthropically-favorable' radius decreases, yet increases the number 
density of stars.  The net effect is a constraint against decreasing 
$\Mp$.  The explicit constraint implied for $\Mp$ is given by
\bea
\left(\frac{1}{6}\hMp^{-\beta}+\frac{5}{6}\hMp^{-1}\right)^{\!2/3}\!
\hMp^{3+4\alpha+6\beta}\mu^{-2/3}s^{-4} \gtrsim 10^{-5}\,, 
\eea 
where $\hat{\mu}_b$ depends relatively weakly on $\Mp$ and has been 
ignored.  The curves that saturate this inequality are displayed in 
Fig.~\ref{fig1} using the label ``$\gamma^{-1}=\tau_{\rm evol}$.''

\subsection{Summary}
\label{sec:summary}

Let us now summarize the results of the previous sections.  Many 
anthropic constraints depend on the primordial curvature perturbation
$\zeta_{\rm eq}$ and on the baryon to photon ratio $\eta$.  Lacking 
any standard model for the generation of either of these, we write them 
generically as $\zeta_{\rm eq}\approx\hMp^{-\alpha}$ and 
$\eta\approx\hMp^{-\beta}$, where $\hMp$ is the ratio between the 
apparent Planck mass and the value obtained within our universe.  For 
the most popular models of inflation, $\alpha$ ranges between one 
($m_\varphi^2\varphi^2$ chaotic inflation) and three (hybrid and natural 
inflation).  Meanwhile, popular models of baryogenesis give $\beta$ 
between zero (efficient leptogenesis and electroweak baryogenesis) and 
3/2 (specific models of SUSY Affleck-Dine baryogenesis). 

Most of the anthropic constraints under consideration are displayed
in Fig.~\ref{fig1}, for representative values of $\alpha$ and $\beta$.  
We assume a WIMP candidate to dominate the dark matter density.  Many 
constraints depend on the total mass within the galactic halo for which 
they are evaluated.  This mass is denoted $\mu$ and is measured in units 
of the Milky Way mass, or $10^{12}$ solar masses.  Note the empty circle 
in each panel of Fig.~\ref{fig1}.  This corresponds to a mass scale equal 
to the mass of the Milky Way with a Planck mass equal to the value 
obtained within our universe.  The region within the $\mu$-$\hMp$ plane 
that is excluded by any constraint is the region that does not include 
this circle.  For clarity we do not display the constraints that 
primordial black holes form a sub-dominant contribution to the energy 
density of the universe and that the dark matter dominates over baryonic 
matter.  These are weaker constraints than those displayed in 
Fig.~\ref{fig1} and they are easy to calculate from Eq.~(\ref{nopbh}) 
and Eqs.~(\ref{baryonfraction}).  Finally, we note that many of the 
constraints in Fig.~\ref{fig1} are deduced by assuming that other 
constraints are satisfied.  For example, the ``disk inst.'' curve is
changed when the constraint represented by the curve 
``$t_{\rm rec}=e^{-2/3}t_{\rm vir}$'' is not satisfied.  The continuous 
curves in Fig.~\ref{fig1} are intended to guide the eye.     

We annotate Fig.~\ref{fig1} as follows.  A number of constraints come 
from the various levels of structure formation.  The curve labeled 
``$\tau_{\rm cool}=\tau_{\rm grav}$'' (this curve has a distinctive 
``dorsal fin'' shape) marks the separation between the mass scales of 
halos that contain baryons which cool faster than they (would) collapse 
and those that do not.  As explained in Section~\ref{sec:cooling}, this 
is one among a set of sufficient, but perhaps not necessary, conditions 
that allow for galaxy formation.  Another one of these conditions is that 
galactic disks be Jeans-unstable, which occurs below the curve labeled 
``disk inst.'' in Fig.~\ref{fig1}.  Meanwhile, structure formation 
requires that over-densities separate from the cosmic expansion before 
the domination of the cosmological constant halts their growth.  This 
requirement is filled below the curve labeled 
``$\sigma_\infty=1.69$.''  Finally, our analysis of structure 
formation assumes that galactic halos virialize after recombination, 
which occurs for $\Mp$ values located above the curve labeled 
``$t_{\rm rec}=e^{-2/3}t_{\rm vir}$.''  Alternative paths to structure
formation are discussed in Appendix~\ref{sec:nonstandard}.

We also consider a few anthropic criteria that are not directly related 
to structure formation.  For example, another constraint that we consider 
is that galaxies not be so small that they are blown apart by 
internal supernovae.  This will not happen if the binding energy of a 
galactic halo well exceeds the energy released via supernovae.  This 
condition is satisfied for $\mu$ and $\hMp$ to the right of the curve 
labeled ``$E_{\rm bind}=E_{\rm sn}$.''  In addition, one might require
that collisions between stellar systems be such that impact parameters 
so small as to dislodge a habitable planet occur on a timescale that is
much larger than the evolutionary timescale, here taken to be 
$\tau_{\rm evol}\approx 5\times 10^9$ yrs.  This constraint is 
satisfied above the curve labeled ``$\gamma^{-1}=\tau_{\rm evol}$.''  
Finally, one might wish to restrict attention to universes that contain 
stars that have surface temperatures in excess of about 3500 K and that 
have main-sequence lifetimes in excess of about four billion years.  
These correspond to positions above the line labeled 
``$\tau_\star = \tau_{\rm evol}$'' in Fig.~\ref{fig1}.     

Except for the stellar lifetime constraint, every constraint displayed in 
Fig.~\ref{fig1} depends on the size of the initial over-density that 
eventually grows into a galaxy.  The curves in Fig.~\ref{fig1} 
correspond to choosing this initial fluctuation to be the rms of the 
density perturbations at a scale $\mu$ evaluated at matter-radiation 
equality.  However, the initial over-density describing any galaxy is a 
stochastic variable that may be larger or smaller than this.  Therefore
all of the curves in Fig.~\ref{fig1} will be shifted when one considers 
galaxies that are away from the norm.  In addition, the disk instability
and close encounters curves (labeled ``disk inst.'' and 
``$\gamma^{-1}=\tau_{\rm evol}$'' respectively) depends very strongly
on other stochastic quantities, such as the galactic spin parameter
(see Section~\ref{sec:closeencounters}).  Therefore the range of $\Mp$ 
that is consistent with the above constraints is larger than the windows 
in Fig.~\ref{fig1} would suggest if one allows observers to arise in 
atypical environments within any given universe.

\section{The Probability Distribution for the Scale of Gravity}
\label{sec:prior}

If the Planck mass $\Mp$ scans across a landscape of universes, then the 
value within any particular universe may not be uniquely determined.  
However, with an understanding of the landscape and a calculus 
to regulate over a conceivably infinite number of infinitely expansive 
universes, we may in principle calculate the distribution of $\Mp$.  
Since we cannot access any of the other universes within the landscape,
such a distribution cannot be directly tested.  Nevertheless, we may 
still use this distribution to calculate the likelihood that we should 
observe the value of $\Mp$ that we do.  As we are forced to test this 
distribution using only our universe, we must be careful to account for 
any selection effects that would attenuate the distribution of $\Mp$.    

These selection effects generate a factor $\mS$ that multiplies the 
``prior'' distribution $\mI$.  Thus we write the probability to measure 
the Planck mass to be $\Mp$, 
\bea
P(\Mp) = \mS(\Mp)\mI(\Mp) \,.
\eea
The factor $\mI(\Mp)$ may be taken as the likelihood for universes with
Planck mass $\Mp$ to arise within the multiverse, while $\mS(\Mp)$ may 
be understood as the likelihood for observers to arise within those 
universes.  We restrict our prior $\mI$ to account for only universes 
exactly like ours except for their value of $\Mp$.  This is equivalent 
to restricting the selection criteria in $\mS$.  As mentioned in the 
introduction, there are many subtle issues that complicate the 
calculation of $\mS$ and $\mI$.  Our purpose here is not to resolve any 
of these issues based on technical grounds.  Instead we explore an 
empirical constraint that may complicate some proposals to address them.

One might expect the likelihood for a universe to support observers to 
be proportional to the total baryonic mass within galaxies in that 
universe.  (Note that here and below we do not presume proportionality
factors to be independent of $\Mp$.)  
Meanwhile, the baryonic mass within 
galaxies is proportional to the total energy within a universe.  This 
quantity diverges in proportion to the volume of the universe.  
Nevertheless, we may hope for a regularization scheme that allows for 
the volumes of universes to be compared.  Since the energy density does 
not redshift during inflation, it is possible that when volumes are 
properly regulated, the ratio between the total energy densities of 
two universes will be proportional to the ratio of their inflationary 
expansion factors.  

While this argument may be intuitively appealing, such a prescription 
for volume-based weighting presents well-known difficulties~\cite{probs3}. 
For instance, its conclusion is crucially dependent on a specific global 
space-like slicing, which is ambiguous outside the horizon of any one 
observer.  For example, an observer can chose a space-like slicing that
is engineered to create a very large initial volume for the observer's 
own universe, while also suppressing the initial volume of the universes 
of casually disconnected observers.  The suppression of an initial volume
can be used to cancel the inflationary expansion factor, such that this
slicing would give a dramatically different counting than the weighting
described above. 

Nevertheless, in at least one proposal this ambiguity has been overcome 
and the result includes a selection effect that weights universes 
according to their inflationary volume (see Garriga {\em et al.} (2006) 
among refs.~\cite{probs3}).  We assume that this result holds and write,
\bea
\mS(\Mp) = \mA(\Mp)\mV(\Mp) \,.
\label{Sdef}
\eea
Here $\mA$ is proportional to the anthropic factor, which ultimately 
gives the likelihood per unit volume for some class of observer to arise 
within a universe.  Depending on one's notion of an observer, $\mA$ 
might include, for example, the baryonic mass fraction within galaxies, 
the fraction of stars with lifetimes in excess of few billion years, etc.  
The factor $\mV$ is the inflationary expansion factor for the universe, 
\bea
\mV(\Mp) = e^{3N(\Mp)} \,,
\label{vfactor}
\eea
where $N(\Mp)$ is the number of e-folds if inflation that typically occurs 
after a universe with Planck mass $\Mp$ has arisen within the multiverse.

If inflation is driven by a single canonical scalar field, then the number 
of e-folds of inflation is 
\bea
N = \frac{1}{\Mp^2}\int_{\varphi_f}^{\varphi_i}\!
\frac{V}{V_\varphi}d\varphi \,.
\label{Ncan}
\eea
Here $\varphi_f$ is the value of the inflaton when inflation ends, set
by when the first slow-roll parameter equals unity, and $\varphi_i$ is 
the value of the inflaton when inflation begins.  Note that in general 
$N$ depends explicitly on $\Mp$.  As a specific example, consider 
chaotic inflation with an inflaton potential 
$V(\varphi)=\frac{1}{2}m_\varphi^2\varphi^2$.  This gives 
\bea
N=\frac{1}{4}\frac{\varphi_i^2}{\Mp^2}-\frac{1}{2} \,.
\label{Ngen}
\eea
It seems evident that $N$ will generically depend on $\Mp$. Yet without 
understanding the mechanism by which a universe is obtained within 
the multiverse, it is not clear what is the (typical) value of 
$\varphi_i$ and what is its dependence on $\Mp$.

To illustrate that the explicit and implicit dependences of $N$ on $\Mp$ 
are not expected to cancel, three models to determine $\varphi_i$ are now 
considered.  The first model sets $\varphi_i$ to be the value where 
classical evolution of $\varphi$ begins to dominate over the quantum 
fluctuations experienced whenever a mode exits the Hubble radius.  This 
is set by the solution to 
\bea
\frac{1}{\sqrt{12}\,\pi}\frac{V^{3/2}}{\Mp^3V_\varphi}= 1 \,.
\label{begininf}
\eea
If this is the case, the total number of e-folds is 
\bea
N \simeq \sqrt{6}\,\pi\frac{\Mp}{m_\varphi} \,.
\label{quantN}
\eea
The second model assumes that $\varphi_i$ is determined by where the 
inflaton energy density equals the Planck energy density.  In this case, 
\bea
N \simeq \frac{\Mp^2}{m_\varphi^2}  \,.
\label{begininf2}
\eea 
On the other hand, if $\varphi_i$ is determined by where the inflaton 
energy density equals $M^4$, then 
\bea
N \simeq \frac{M^4}{m_\varphi^2 \Mp^2}  \,.
\eea
Not only does $N$ generally depend on $\Mp$, but the dependence is very
strong for $\Mp$ near the value obtained within our universe.  Consider 
for example the case of chaotic inflation with $\varphi_i$ set by 
Eq.~(\ref{begininf}).  Then $N\sim 10^5\,\hMp$, where $\hMp$ is the Planck 
mass in units of the value obtained within our universe.  Meanwhile, if 
Eq.~(\ref{begininf2}) sets the value of $\varphi_i$, then 
$N\sim 10^{10}\,\hMp^2$.  Clearly different choices for $\varphi_i$, and in 
particular different models of inflation, will in general give a different 
dependence of $N$ on $\Mp$.  However, the dependence is always strong.  
This is because our universe experienced a large number (at least about 
sixty) of e-folds of inflation.  

The ambiguity over the $\Mp$ dependence of $N$ is not of concern.  The
important result is that so long as the dependence on $\Mp$ of $\mA$ 
and $\mI$ is significantly weaker than the strong exponential dependence
in $\mV$, then we expect $\Mp$ to be most probably observed very near 
one of the boundaries of the anthropic range.  The analysis of   
Section~\ref{sec:anthropic} reveals that this is not the case with at 
least the factor $\mA$ and the value of $\Mp$ observed within our universe.  
We illustrate this with an explicit example in Appendix~\ref{sec:MpLambda}.  
There we show that even in the contrived case where $N\sim 60\,\hMp$, the 
volume factor $\mV(\Mp)$ overwhelms what appears to be one of the tightest 
anthropic constraints.  This pushes the expectation value for $\Mp$ well 
beyond what we estimated to be the anthropic boundary, while the value 
obtained within our universe sits far down the tail of the distribution.  
This is exactly analogous to the runaway ``$\sigma$-problem'' and the 
``$Q$ catastrophe'' introduced in refs.~\cite{FHW}.  We refer to our 
example as the ``$\Mp$-problem.''    

The $\sigma$-problem and $Q$ catastrophe were motivated by the fact that 
in many models of inflation the total number of e-folds of inflation 
depends on the inflationary parameters that also set the level of 
density perturbations (the authors of refs.~\cite{FHW} use the notations 
$\sigma\sim Q\sim\zeta$).  For example, in chaotic inflation with 
potential $V(\varphi)=\frac{1}{2}m_\varphi^2\varphi^2$ one finds 
$N\sim\zeta^{-1}$.  Therefore if the inflationary parameters may scan
over the landscape, by the same argument given above we expect $\zeta$
to be pushed to one of its anthropic boundaries, whereas in our universe
it sits comfortably near the middle of the anthropic 
window~\cite{TARW}.  It has been pointed out~\cite{Linde:2005yw} that 
this argument is not completely satisfactory, since by hypothesis 
universes with an enormous number of e-folds are preferred.  In such 
universes, $\zeta$ may plausibly depend on different parameters during
a long stretch of early inflation than it does near the end of inflation, 
when scales important to the formation of structure are generated.  
Moreover, we note that the curvature perturbation is related to the 
first slow-roll parameter, 
\bea
\zeta\sim \frac{1}{\sqrt{\epsilon_I}}\frac{V^{1/2}}{\Mp^2} \,.
\eea
Inflation of longer duration requires a smaller $\epsilon_I$, yet for 
inflation to end at all requires that at some point $\epsilon_I$ evolve 
toward unity.  Therefore $\zeta_{\rm eq}$ may be significantly decreased 
from its value during most of inflation by the necessary condition that 
$\epsilon_I$ interpolate between some very small value and unity by the 
end of inflation.  

We emphasize that the $\Mp$-problem is not hampered by these issues.  
That is, unlike $\zeta$, $\Mp$ is a constant within any given 
universe.\footnote{Of course, the inflationary landscape hypothesis 
presumes that the fields $\phi$ described in the introduction, 
c.f. Eq.~(\ref{lel}), will evolve within the multiverse.  Indeed, 
this is how the landscape is populated.  However, we assume that the 
vacua defined by the fields $\phi$ are selected prior to period of 
inflation in which we take interest, during which $\Mp$ is constant.  
Specifically, the factor $\mI$ is assumed to account for any selection 
effects due to the field evolution prior to this period of inflation, 
and the terms $\mA$ and $\mV$ are defined to apply only after a 
particular meta-stable state, with a specific value of $\Mp$, has been 
selected.}  
We also emphasize that if the model of inflation that describes our
universe exhibits an $\Mp$-problem, then allowing more parameters to 
vary across the landscape cannot mitigate this problem.  That is, 
although allowing more parameters to vary might dramatically shift the 
expectation value for $\Mp$ after the additional parameters have been 
marginalized, this can only happen if the overwhelming majority of 
universes near the new expectation value have values for the other 
parameters that are very different than ours.  We would still be left 
with the challenge to explain why we find ourselves in a universe like 
ours, and not with these different parameter values.

There are significant caveats to this result.  First of all, it is not 
clear that the selection effects in $\mS$ should actually factorize
as in Eq.~(\ref{Sdef}).  Since the diverging volumes of sub-universes is 
one of the circumstances that complicates making landscape predictions, 
we cannot be assured that the resolution of this problem will result in 
universes with greater inflationary expansion factors being more likely 
to harbor observers.  Another caveat to this discussion is that little 
is known about the distribution $\mI(\Mp)$.  As we have defined it, this
term receives two separate contributions.  One contribution comes from
the distribution of $\Mp$ values over the landscape, that is the 
frequency of $\Mp$ values among the number of meta-stable states that are
allowed by the underlying theory.  A second contribution comes from the 
dynamics of the multiverse, which may prefer certain meta-stable states 
over others as the multiverse evolves in time.  This is because the 
tunneling and diffusion rates of quantum fields will in general depend 
on $\Mp$, such that meta-stable states with certain values of $\Mp$ 
will appear more frequently within the multiverse than others.  This
$\Mp$ dependence within $\mI(\Mp)$ could be very strong; see for example
the studies of quantum diffusion in refs.~\cite{probs2}.

Therefore $\mI(\Mp)$ may depend more sharply on $\Mp$ than does 
$\mV(\Mp)$, with a local peak within the anthropic range.  This might at 
first seem incredibly fortuitous.  However, the situation is very 
different from the case of the cosmological constant $\rho_\Lambda$.  In 
that case we observe $\rho_\Lambda$ to be very far from its `natural' 
value and therefore we must presume a very diverse and densely packed 
landscape in order for the value that we observe to exist at all.  However, 
since we do not know the natural value of $\Mp$, its landscape window 
could be much smaller.\footnote{It is tantalizing that within the context 
of weighting universes by inflationary expansion factors, chaotic 
inflation with $N$ set by either Eq.~(\ref{quantN}) or 
Eq.~(\ref{begininf2}) pushes $\Mp$ to larger values, while the largeness 
of $\Mp$ relative to other mass scales is well-noted in our universe.  Let 
us assume a fundamental scale $M\sim M_{\rm GUT}$. Then in this case we 
expect $\Mp\gg M_{\rm GUT}$, and it is possible that 
$\Mp\sim 10^3M_{\rm GUT}$ is simply the largest that the landscape allows.  
Furthermore, Eq.~(\ref{quantN}) pushes the inflaton mass $m_\varphi$ to 
smaller values, and perhaps $m_\varphi$ is the smallest that the landscape
allows.  Thus we obtain the apparent hierarchy 
$\Mp\gg M_{\rm GUT}\gg m_\varphi$.}
In addition, it is possible that the landscape is not as densely populated 
as we have presumed, in particular once we restrict attention to 
meta-stable states in every way like ours except in the value of $\Mp$.  
For example, if the spacings between allowed values of $\Mp$ are 
significant next to the size of the anthropic window, then our value of 
$\Mp$ might be consistent with the shape of $\mI(\Mp)$.

Furthermore, the model of inflation that describes our universe may not 
actually exhibit an $\Mp$-problem.  This would happen, for example, if the 
number of e-folds of inflation that describe this model were independent 
of $\Mp$ or had a maximum for some finite value of $\Mp$.  An interesting
example of the latter case occurs when the effective Planck mass is not 
fixed within our meta-stable state, but evolves as in Brans-Dicke theory.  
This scenario has been studied in refs.~\cite{probs2}, where it is shown 
that for some non-minimally coupled models of inflation, the inflationary 
expansion factor is maximized when inflation ends at some finite value of 
$\Mp$.  Note however that for this or any other model of inflation to 
avoid the $\Mp$-problem, it would have to generate more e-folds of 
inflation than all of the other anthropically viable possibilities within 
the landscape.  Moreover, the value of $\Mp$ that maximizes $N$ would have 
to lie within the anthropic window.  

Finally, it is possible that the analysis of Section~\ref{sec:anthropic}
missed or underestimated an important anthropic condition.  This might 
appear as the most attractive possibility, but one must be careful to 
appreciate the strength of the exponential dependence within $\mV(\Mp)$.  
In order to cancel this exponential dependence and thus make the observed
value of $\Mp$ reasonably likely, an anthropic constraint must appear to 
exponentially suppress the likelihood for observers to arise within our
universe.  The observed prevalence of galaxies, long-lived stars, 
supernovae, and planets, along with the observation that our solar system 
does not seem to occupy a particularly over- or under-dense region within 
the Milky Way, all seem to suggest that this is not the case.  Since 
there do not yet exist experimentally confirmed theories for inflation, 
reheating, and baryogenesis, it is still possible that one of these 
processes presents an anthropic selection effect that provides this 
exponential suppression. This possibility is explored relative to 
the reheating temperature and baryogenesis in the context of 
the $\sigma$-problem in ref.~\cite{Hall:2006ff}.  Since the reheating 
temperature in general also depends on $\Mp$, this analysis applies 
equally to our scenario.

\section{Anthropic Constraints on $\Lambda$ and the Scale of Gravity}
\label{sec:cc}

It is straightforward to extend the analysis of 
Section~\ref{sec:anthropic} to the case where both $\Mp$ and the
cosmological constant $\rho_\Lambda$ may (independently) scan over a 
landscape.  The only constraint that is affected by this generalization 
is the requirement that over-densities separate from the Hubble flow 
before their growth is halted by the domination of the cosmological 
constant.  The maximum amplitude reached by a linear rms over-density 
in this scenario is,
\bea
\sigma_\infty \approx 1.44\times\frac{3}{5}
\frac{a_\Lambda}{a_{\rm eq}}\sigma_{\rm eq}
\approx 3.20\,\hat{\rho}_{\rm eq}^{1/3}\hat{\rho}_\Lambda^{-1/3}
\hat{\sigma}_{\rm eq} \,.
\label{sigmainf} 
\eea
On the other hand, an over-density has separated from the Hubble flow 
when a linear analysis gives $\sigma\geq 1.69$~\cite{virialize}.  
Therefore an rms fluctuation will eventually form a halo if 
$\sigma_\infty\geq 1.69$, which gives the generalization of 
Eq.~(\ref{lambdaconstraint1}):
\bea
\hat{\rho}_{\rm eq}\hat{\sigma}_{\rm eq}^{3}\hat{\rho}_\Lambda^{-1}
\gtrsim 0.1 \,,
\label{lambdaconstraint2}
\eea
where we find it convenient to henceforth use $\hat{\sigma}_{\rm eq}$ 
instead of $\hat{\zeta}_{\rm eq}s$.  This is the only result from 
Section~\ref{sec:anthropic} that changes when $\rho_\Lambda$ may
scan over the landscape.    
  
Clearly, Eq.~(\ref{lambdaconstraint2}) is weakened as $\rho_\Lambda$ is 
decreased from the value it obtains within our universe.  In this case, 
Eq.~(\ref{lambdaconstraint1}) eventually ceases to be the strongest 
constraint and $\Mp$ is bounded from above by one of the other curves 
in Fig.~\ref{fig1}.  We may also interpret Eq.~(\ref{lambdaconstraint2}) 
as an upper bound on $\rho_\Lambda$ for a specified value of $\Mp$.  
In universes with a larger value of $\Mp$, $\rho_\Lambda$ is then more 
tightly bound than in our universe.  However, in universes where $\Mp$ 
is smaller than in our universe, the bound on $\rho_\Lambda$ may be 
significantly weakened.  This effect can be dramatic.  For example, if 
we take $\hMp=0.1$ and if $\alpha=1$ and $\beta=0$, then $\rho_\Lambda$ 
may be increased by a roughly a factor of ten million and still satisfy 
Eq.~(\ref{lambdaconstraint2}).  Of course, to determine the most likely 
range within which to observe $\rho_\Lambda$ requires to determine the 
prior distribution $\mI(\rho_\Lambda,\Mp)$ and to incorporate all of the 
selection effects into a factor $\mS(\rho_\Lambda,\Mp)$, as described in 
Section~\ref{sec:prior}.  Both of these tasks are beyond the scope 
of this paper.  

Nevertheless, it is worthwhile to proceed but within a very simplified 
picture.  While our level of analysis does not permit even an 
approximate landscape prediction, our results do imply restrictions on
the dependence of $\mS$ and $\mI$ on $\Mp$.  Our first assumption is 
that the landscape is so densely packed that we can approximate the prior 
distribution $\mI(\Mp,\rho_\Lambda)$ to be a continuous and smooth 
function of both $\Mp$ and $\rho_\Lambda$.  Then we can write the 
probability distribution for $\rho_\Lambda$ in the form
\bea
P(\rho_\Lambda)\propto\int\!\mS(\Mp,\rho_\Lambda)
\mI(\Mp,\rho_\Lambda)d\Mp \,.
\label{fullprob}
\eea
We discuss in Section~\ref{sec:prior} and in Appendix~\ref{sec:MpLambda} 
how our universe appears extremely unlikely to be observed if $\mS$ 
contains a factor proportional to the inflationary expansion factor.  
Since we wish to expose additional restrictions on $\mS$ and $\mI$, we 
now assume that $\mS$ does not contain this factor.

It is helpful to first consider the distribution $P(\rho_\Lambda)$ when
$\Mp$ is fixed to the value obtained within our universe.  This 
corresponds to taking $\mS\propto\delta(\hMp-1)$ and thus eliminating
the integral in Eq.~(\ref{fullprob}).  Refs.~\cite{anthcc} argue that it 
is appropriate to restrict attention to only positive values of 
$\rho_\Lambda$ and to take the distribution $\mI$ to be roughly 
independent of $\rho_\Lambda$.  Although the rms fluctuation 
$\sigma_{\rm eq}$ is constrained by Eq.~(\ref{lambdaconstraint2}), any
particular over-density may be larger or smaller than $\sigma_{\rm eq}$.
This implies that galaxies of a given mass will form in universes
even when $\rho_\Lambda$ is larger than what is allowed by 
Eq.~(\ref{lambdaconstraint2}).  On the other hand, galaxies of a given 
mass become statistically rarer as $\rho_\Lambda$ is increased.  To 
account for this, it is customary to speculate that the likelihood for 
a particular universe to be observed is proportional to the fraction of
its total mass that collapses into galaxies with masses above some
minimum $\mu_{\rm min}$~\cite{anthcc}.  This minimum galaxy mass is 
presumably set by other anthropic considerations.    

The spectrum of density perturbations is at least approximately described
by Gaussian statistics.  Therefore a randomly selected co-moving volume 
may or may not collapse, depending on the size of the matter over-density
contained within the volume.  We parameterize volumes using the mass $\mu$
that they enclose, measured in units of the Milky Mass, $10^{12}M_\odot$. 
Then the likelihood that a mass $\mu$ will eventually separate from the 
cosmic expansion is given by the Press-Schechter function~\cite{virialize},
\bea
F(\mu) \!&=&\! \sqrt{\frac{2}{\pi}}\frac{1}{\sigma_\infty(\mu)}
\int_{1.69}^\infty\!
\exp\left[-\frac{1}{2}\frac{z^2}{\sigma_\infty^2(\mu)}\right] dz 
\phantom{\Bigg(\Bigg)}\nn\\
\!&=&\! {\rm erfc} \left[ \frac{0.373\,\hat{\rho}_\Lambda^{1/3}}
{\hat{\rho}_{\rm eq}^{1/3}\hat{\sigma}_{\rm eq}(\mu)}\right] \,.
\label{psfunction}
\eea
The percentage of over-densities that eventually virialize is a function
of the enclosed mass $\mu$ because the rms amplitude of the initial 
density perturbations $\sigma_{\rm eq}$ depends on $\mu$ (see 
Section~\ref{sec:structure}).  The fraction of galaxies that have mass 
between $\mu$ and $\mu+d\mu$ is $(dF/d\mu)d\mu$.  Since $F(\mu\to\infty)=0$, 
this means that the fraction of mass contained within galaxies with mass 
above the mass scale $\mu$ is simply the Press-Schechter function $F$ 
evaluated at $\mu$.  When only the cosmological constant scans over the 
landscape, $\hat{\rho}_{\rm eq}=\hat{\sigma}_{\rm eq}=1$.  This gives  
$P(\rho_\Lambda)\propto\mS(\rho_\Lambda)\propto 
F(\mu_{\rm min},\rho_\Lambda)$~\cite{anthcc,TARW}.

In order to study the scenario where both $\rho_\Lambda$ and $\Mp$ scan 
over the landscape, we adopt a very simplified picture.  First, we assume 
that $\mI$ is independent of both $\rho_\Lambda$ and $\Mp$ over the 
anthropically allowed window.  We emphasize that, unlike the case with 
$\rho_\Lambda$, we are unaware of any physical justification for this 
assumption regarding $\Mp$.  Second, we restrict our attention to galaxies
with masses near the mass of the Milky Way.  We perform this restriction 
simply so that we may ignore the scale dependence of anthropic constraints.  
It turns out that values of $\Mp$ somewhat larger than our own do not 
contribute significantly toward $P(\rho_\Lambda)$.  To highlight this 
result we simply neglect all constraints on increasing $\Mp$.  On the 
other hand, the selection effects that bound $\Mp$ from below are very 
important when determining $P(\rho_\Lambda)$.  For simplicity we
consider selection effects from only one additional constraint; which is
that stellar encounters are rare enough on average to allow for life to 
evolve in the intervening time.  According to Fig.~\ref{fig1}, this is 
usually the strongest constraint on decreasing $\Mp$.  The exception 
appears to be the case of low $\alpha$ and low $\beta$, where the galactic 
cooling constraint can interfere and the stellar lifetime constraint is 
not far below the close encounters constraint.  We simply ignore the 
cooling constraint and note that we could just as well evaluate 
$P(\rho_\Lambda)$ for galaxy masses somewhat below the mass of the Milky 
Way to obtain a similar result.  To account for stellar lifetimes, we 
impose a hard cut-off below $\hMp=0.1$.

As mentioned above, the mass fraction within galaxies with masses between 
$\mu$ and $\mu+d\mu$ is $\delta F\equiv (dF/d\mu)d\mu$. 
This quantity in 
general depends on the time at which one looks at the universe.  We 
count galaxies in the infinite future, which is practically equivalent to
counting galaxies at any time after the domination of $\rho_\Lambda$.
Then \bea
\frac{d F}{d \mu} \propto \frac{1}{\sigma_\infty^2}\, 
\bigg|\frac{d \sigma_\infty}{d \mu}\bigg|\,
e^{-1.43/\sigma_\infty^2} \,.
\eea
The only $\mu$ dependence within $F$ stems from the dependence on 
$\sigma_\infty\propto\sigma_{\rm eq}\propto s(\mu)$, where $s(\mu)$ is 
given by Eq.~(\ref{sdef}).  
Within any given universe, 
to consider only galaxies with a particular mass $\mu$ in the far future 
is equivalent to selecting only over-densities with a particular amplitude 
at equality.  This is because within that universe over-densities with 
smaller amplitudes will form galaxies with smaller mass while 
over-densities with larger amplitudes will form galaxies with larger mass 
(recall that we look at the universe after $\rho_\Lambda$ domination when 
the growth in over-densities has halted).  The amplitude of the initial 
over-density that is selected by looking at a particular galaxy mass 
$\mu$ is the one that gives $\sigma(\mu)= 1.69$ in the infinite future.  

We must now account for the close encounter constraint mentioned above.  
This constraint is converted into a selection effect by noting that if 
the rate of disastrous encounters between stellar systems is $\gamma$, 
then the probability that a stellar system will survive for a time 
$\tau$ is $e^{-\gamma\tau}$.  The rate $\gamma$ is discussed in 
Section~\ref{sec:closeencounters}.  Note that it depends on the 
amplitude of the initial over-density that seeded the galaxy.  We 
restrict our attention to galaxies with masses near to the mass of the 
Milky Way.  As described above, these galaxies only come from 
over-densities that satisfy $\sigma(\mu\approx 1)=1.69$ in the infinite
future.  At equality, these over-densities have an amplitude
\bea
\sigma\approx  5\times 10^{-4}\,\hat{\rho}_\Lambda^{1/3}
\hat{\rho}_{\rm eq}^{-1/3} \,.
\eea
Now we can write the likelihood $P_{\rm ss}$ that a stellar system will 
survive for at least a time $\tau$ within this set of galaxies.  We take 
$\tau=\tau_{\rm evol}\approx 5\times 10^{9}$ yrs, which gives
\bea
P_{\rm ss}
\approx \exp\left( -7\times 10^{-7}\hMp^{-3}\hat{\mu}_b^{1/2}
\hat{f}_b^6\,\hat{\rho}_\Lambda^{\,4/3} \mu^{2/3} \right) \,,
\eea 
where the dependence on a general mass scale $\mu$ has been 
restored for future reference. 

\begin{figure}[t!]
\includegraphics[width=8.75cm]{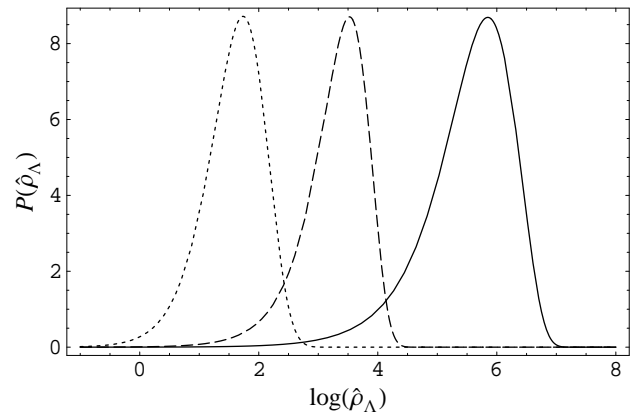}
\caption{The distribution $P(\hat{\rho}_\Lambda)$ displayed against 
$\log (\hat{\rho}_\Lambda)$.  The solid curve is $P(\hat{\rho}_\Lambda)$ 
marginalized over universes in which $\Mp$ may vary and with $\alpha=1$ 
and $\beta=0$, the longer-dashed curve is the same quantity but for 
$\alpha=3$ and $\beta=3/2$, while the shorter-dashed curve is 
$P(\hat{\rho}_\Lambda)$ evaluated when $\Mp$ is fixed to the value within 
our universe. All three distributions 
are for fixed galactic masses $\mu=1$. See 
text for results obtained for a range of galactic masses. 
The normalizations are chosen for clarity.}
\label{fig3}
\end{figure}

So far our assumptions correspond to weighting universes by the fraction 
of stellar systems that survive close encounters for longer than 
$\tau_{\rm evol}$ and that exist in galaxies with mass near to the Milky
Way mass.  We also require $\Mp\geq 0.1$ in order to ensure that 
sufficiently long-lived stars exist in these universes.  Finally, we 
should account for the fact that the abundance of baryons relative dark
matter will depend on the value of $\Mp$ within each universe.  Putting
all of this together gives the probability density,
\bea
P(\hat{\rho}_\Lambda) \!&\propto&\! \int^{\mu_{\rm max}}_{\mu_{\rm min}} 
\!\!d\mu \int_{0.1}^\infty \!d\Mp f_b(\Mp)
P_{\rm ss}(\Mp,\hat{\rho}_\Lambda,\mu) 
\nonumber \\ 
& & \times \frac{d}{d \mu} F(\Mp,\hat{\rho}_\Lambda,\mu)
  \,.\,\,\,
\label{simpprob}
\eea  
The full $\Mp$ dependence of $f_b$, $P_{\rm ss}$, and $d F/ d \mu$ is 
found by substitution of the results from Section~\ref{sec:anthropic}.
In general Eq.~(\ref{simpprob}) is integrated over a window 
$\mu _{\rm min} \leq \mu \leq  \mu_{\rm max} $, but as 
motivated above, in our main analysis we restrict to a 
narrow window about $\mu=1$.
Finally, as explained previously, Eq.~(\ref{simpprob}) makes the simplifying 
but unrealistic assumption that $\mI(\Mp,\rho_\Lambda)\approx$ constant.   

The results of a numerical computation of $P(\rho_\Lambda)$ are displayed 
in Fig.~\ref{fig3}.  For reference, we also display the result when $\Mp$ 
is fixed to the value obtained within our universe (note that this 
corresponds to $dF/d\mu|_{\mu=1}$ and not $F(\mu=1)$).  Our value of 
$\rho_\Lambda$ corresponds to the origin on this graph.  Evidently the 
assumptions of this section render the observation of $\rho_\Lambda$ at 
or below our value very unlikely. In fact, the fraction of 
$P(\rho_\Lambda)$ that sits below $\hat{\rho}_\Lambda=1$ is about
$7\times 10^{-5}$ for $\alpha=1$, $\beta=0$ and about $4\times 10^{-4}$
for $\alpha=3$, $\beta=3/2$. Since relatively large values of 
$\rho_\Lambda$ receive significant weight only when $\Mp$ is relatively 
small, we also see that most of the weight of these distributions comes 
from values of $\Mp$ that are below the value obtained within our universe.  
This is evidence of a sort of `statistical pressure' that gives greater 
weight to those values of marginalized parameters that permit a larger 
value of $\rho_\Lambda$.  This is why it was unimportant to account for 
selection effects that constrain $\Mp$ from above. 

Here we note the importance of the close encounters constraint and of 
recombination timing constraint (i.e. the 
``$t_{\rm rec}=e^{-2/3}t_{\rm vir}$'' constraint) in bounding the 
anthropically allowed variation of $\hat{\rho}_{\Lambda}$.  Inspecting 
Fig.~\ref{fig1} indicates that of the constraints limiting $\Mp$ from 
below, the close encounters bound is the strongest. For $\mu=1$, it 
bounds $\hMp\gtrsim 10^{-0.2}$ for $\alpha=3$ and $\beta =3/2$, and 
approximately $\hMp \gtrsim 10^{-0.8}$ for $\alpha=1,\beta =0$. 
Inserting these limits on $\hMp$ into Eq.~(\ref{lambdaconstraint2}) 
gives maximum values for $\hat{\rho}_{\Lambda}$ in good agreement with 
the peaks in Fig.~\ref{fig3}. Had only the star lifetime constraint 
$\hMp\gtrsim 0.1$ been imposed, the maximum allowed value of 
$\hat{\rho}_{\Lambda}$ would have been much larger. For 
$\alpha=1$ and $\beta=0$, $\hat{\rho}_{\Lambda}$ could be as large as 
$4\times10^7$. In the case of $\alpha=3$ and $\beta=3/2$, 
$\hat{\rho}_{\Lambda}$ can be as large as $2\times 10^{14}$. But the 
anthropic constraints in Fig.~\ref{fig1} for this latter model indicate 
that the ``$t_{\rm rec}=e^{-2/3}t_{\rm vir}$'' constraint is stronger 
than the star lifetime constraint. Imposing the recombination constraint 
requires instead that $\hat{\rho}_{\Lambda}$ be no larger than 
about $4\times 10^8$. 

The analysis that leads to the curves in Fig.~\ref{fig3} gives at best a 
crude approximation for the actual probability distribution for 
$\rho_\Lambda$.  One improvement to the analysis would be to weight 
universes by the mass fraction that collapses into galaxies that have a 
range of anthropically favorable masses, instead of the fraction that 
collapses into only galaxies with the Milky Way mass.  Including galaxies 
with greater masses will tend to push the weight of the distributions 
toward smaller $\rho_\Lambda$, while including galaxies with smaller 
masses pushes the weight of the distributions toward larger 
$\rho_\Lambda$.  We have checked that under the assumptions outlined 
above, allowing for a range of galaxy masses $0.1\leq\mu\leq 10$ tends to 
push the weight of the distributions $P(\rho_\Lambda)$ to slightly larger 
values of $\rho_\Lambda$.  

Previous calculations of the distribution $P(\rho_\Lambda)$ integrate
over all galaxy masses equal to or larger than the Milky Way mass.  
Although our anthropic considerations offer no reason to ignore galaxies
with mass below that of the Milky Way, and galactic cooling constraints 
limit the formation of galaxies with larger masses, we nevertheless 
consider the evaluation of Eq.~(\ref{simpprob}) for a range of masses 
$1 \leq \mu \leq\infty$.  Integrating over $0\leq\hat{\rho}_{\Lambda}\leq 1$ 
gives the probability $P_*$ that observers in such galaxies would observe 
a cosmological constant less than or equal to our own.  Numerically, we 
find $P_*=1\times 10^{-3}$ for $\alpha=3,\beta=3/2$ and 
$P_*=2\times 10^{-4}$ for $\alpha=1, \beta=0$. For comparison, we find 
that in our universe $P_* = 0.06$. To reiterate, these calculations 
ignored any additional selection effects that might depend on $\mu$, 
such as the effects of different galactic cooling rates.

Another improvement to the analysis would be to include more 
$\Mp$-dependent selection effects.  The abundance of heavy elements and
of long-lived stars with appropriate surface temperatures both seem 
important when determining the likelihood for observers to arise within 
a universe.  However, the analysis of Section~\ref{sec:anthropic} does 
not shed light on how to calculate these selection effects.  One thing
that is clear is that the range of typical stellar masses scales as 
$\Mp^3$.  This means that galaxies of a fixed mass will contain more 
stars as $\Mp$ is decreased.  If all else were equal, as $\Mp$ is 
decreased this would result in a greater number of observers per unit 
baryon mass in a galaxy, 
which would tend to push the weight of $P(\rho_\Lambda)$ toward larger 
values of $\rho_\Lambda$.  In addition, the rate of destructive 
encounters $\gamma$ is a function of stochastic variables, including for 
example the spin parameter $\lambda$.  The statistical distribution of 
these variables could tend to strengthen or weaken the close encounter 
constraint as a function of $\Mp$.  However, when everything else is equal
the `statistical pressure' alluded to below Eq.~(\ref{simpprob}) tends to 
give greater weight to those values of stochastic variables that allow for 
a smaller $\Mp$ and larger $\rho_\Lambda$.\footnote{We have confirmed this 
phenomena with the following simple example.  The close encounter rate 
$\gamma$ is proportional to a factor $f_\star$ that accounts for the 
increased density of the galactic disk relative the dark matter halo
(see Section~\ref{sec:starformation}).  This factor depends sensitively
on $\lambda$, $f_\star\propto \lambda^{-8}$.  In the preceding analysis,
we normalized the factor $f_\star$ so as to give the correct stellar 
density within our neighborhood of the Milky Way.  However, $\lambda$ is 
a stochastic variable.  N-body simulations suggest that the distribution 
for $\lambda$ can be approximated using~\cite{Bullock:2000ry}
\bea
P_\lambda(\lambda)d\lambda \propto \frac{d\lambda}{\lambda}
\exp\left[-2\ln^2( 28.6\lambda)\right] \,.
\eea     
This distribution has a peak at about $\lambda\approx 0.03$, while the 
Milky Way appears to be described by $\lambda\approx 0.06$~\cite{MWprop}.  
This implies that typical values of $\lambda$ more tightly constrain 
$\Mp$ than the value represented in Fig.~\ref{fig3}.  Therefore, one might 
expect that when we treat $\lambda$ as a stochastic variable with 
distribution $P_\lambda(\lambda)$, that the weight of the distributions 
$P(\rho_\Lambda)$ will shift to smaller values of $\rho_\Lambda$.  In fact,
the opposite trend occurs, as the previously mentioned ``statistical 
pressure'' is such that the weight of the distributions $P(\rho_\Lambda)$ 
actually shifts the location (in $\hat{\rho}_{\Lambda}$) of the peak  
by approximately an order of magnitude toward larger values of 
$\rho_\Lambda$.}

Of course, a proper calculation of $P(\rho_\Lambda)$ requires an 
understanding of the prior distribution for $\Mp$, $\mI(\Mp)$.  It must
be emphasized that shape of $\mI(\Mp)$ could dramatically influence the 
shape of the distribution $P(\rho_\Lambda)$.  Therefore the results of
this section are best understood as an empirical restriction on the 
dependence of $\mI(\Mp)$ on $\Mp$.  Since this is our main point in
this section, let us be very explicit.  The curves displayed in 
Fig.~\ref{fig3} suggest that within a very simplified landscape picture, 
it is very unlikely to observe a value of $\rho_\Lambda$ that is at or 
below the value within our universe.  This means that if a landscape 
picture is to describe our universe, it should contain important 
ingredients that were neglected in our analysis.  In addition, these 
additional ingredients should provide a strong emphasis for larger 
values of $\Mp$.  Thus we conclude that for a landscape picture to  
describe our universe as among those that are likely to be observed, 
it is necessary that $\mI(\Mp)$ or some other neglected selection effect 
must receive the vast majority of its weight for values of $\Mp$ that 
are very near to or larger than the value obtained within our universe.    
We emphasize that the analysis of this section did not assume that the 
inflationary expansion factor enters into landscape calculations.

\section{Conclusions}
\label{sec:conclusions}

If the magnitude of the apparent Planck mass $\Mp$ may scan across a 
landscape of possibilities, then there may exist universes with 
physical parameters and interactions in every way like those within 
our universe except for their value of $\Mp$.  We have calculated 
the range over which $\Mp$ may scan over such universes while still
satisfying a number of anthropic constraints.  Perhaps not 
surprisingly, if we combine all of the anthropic constraints we find a 
rather narrow window for allowed $\Mp$.  The results for WIMP dark matter
and representative models of inflation and baryogenesis are displayed 
in Fig.~\ref{fig1}.  Of course, the window for allowed $\Mp$ is expanded 
if one loosens the anthropic criteria.  

More interestingly, this window will expand if an important 
cosmological quantity is determined by a stochastic process.  For 
example, many scenarios to generate a primordial curvature 
perturbation depend on the local vev of a light scalar field, as 
does the density of dark matter when it is determined by the axion.  If 
these models apply, then the curvature perturbation and/or dark matter
density are not correlated with changes in $\Mp$, and a much larger
window for $\Mp$ may be able to satisfy anthropic constraints.  Our
purpose has been to calculate a minimal window for allowed $\Mp$, so 
we have not considered these possibilities in detail.

Even a very small window for allowed values of $\Mp$ has important
implications for the landscape paradigm.  In particular, the probability 
to observe a particular value of $\Mp$ may be weighted by the inflationary 
expansion factor of universes that contain that value.  This effect inputs
a strong exponential dependence on $\Mp$ into the probability 
distribution, which must be offset by another strong selection effect near
the peak of the distribution.  This other selection effect could be a 
very sharp peak or boundary to the underlying landscape distribution; 
otherwise the effect must come from an exponentially strong anthropic 
dependence on $\Mp$.  Such a strong anthropic dependence on $\Mp$ 
would be in conflict with the observation that $\Mp$ has even a narrow
anthropic window in our universe.  This is another example of the 
runaway inflation problem discussed in the recent literature.

We also consider the anthropic window for the cosmological constant 
$\Lambda$ when both $\Lambda$ and $\Mp$ are allowed to independently 
scan over the landscape.  Even when the allowed range for $\Mp$ is 
relatively narrow, it still allows for a significant broadening of the 
allowed range for $\Lambda$.  This is because $\Lambda$ is only 
constrained by the necessity that cosmic structures separate from the 
Hubble flow before $\Lambda$ domination.  Meanwhile, the time at which
structures separate from the cosmic expansion is proportional to a 
high power of $\Mp$.  The result is that even for values of $\Mp$ 
within the small allowed windows of Fig.~\ref{fig1}, $\Lambda$ may be 
over ten million times larger in other universes than it is within ours.  
Just because $\Lambda$ may be larger does not automatically imply that 
our value of $\Lambda$ is less likely to be observed, since selection
effects may ultimately weight smaller values of $\Lambda$ more than 
larger values.  We perform a very basic calculation which suggests that
anthropic selection effects tend to make larger values of $\Lambda$ more 
likely to be observed.  This suggests that the observation of a 
cosmological constant at or below the level obtained within our universe 
is very unlikely unless unknown anthropic selection effects or the 
underlying landscape distribution of $\Mp$ is dominated by values very 
near to or larger than the value obtained within our universe.

\begin{acknowledgments}
The authors thank Stearl Phinney, Jonathan Pritchard, Scott Thomas, and 
Andrei Linde for helpful discussions.  This work was supported by the U.S. 
Department of Energy under contract Nos. DE-FG03-92ER40689 and 
DE-FG03-92ER40701.    
\end{acknowledgments}

\appendix
\section{Non-Standard Paths Toward Structure Formation}
\label{sec:nonstandard}

One might wonder what are the constraints on structure formation if 
we do not assume that dark matter dominates over baryonic matter, or that
virialization occurs after recombination.  If dark matter does not 
dominate over baryonic matter, then the evolution of over-densities in 
the dark matter does not significantly affect the evolution of 
over-densities in baryons.  Without appreciable dark matter potential 
wells, baryon over-densities do not grow (even logarithmically) until 
after recombination. This is because in the era before recombination, 
the Jeans length for the tightly coupled baryons, 
\bea 
R_J = \sqrt{{\textstyle \frac{8}{3}}}\pi H^{-1} v_s \,,
\eea
where $v_s= 1 /\sqrt{3}$ is the speed of sound prior to recombination, 
is always larger than the Hubble radius. Growth therefore does not occur 
in either the radiation or baryon-dominated era until after recombination.   
Between recombination and the domination of cosmological constant the 
evolution of over-densities may be approximated by
\bea
\sigma \approx (a/a_{\rm rec})\sigma_{\rm rec} \,.
\eea  
The spectrum of fluctuations at recombination $\sigma_{\rm rec}$ is scale
dependent in the sense that it is constant for scales larger than the 
Hubble radius at recombination but rapidly decreases to zero as one looks 
at smaller distance scales.  This is because of the tight coupling between 
baryon and radiation over-densities, and because the latter decay after 
they enter the Hubble radius.  

As in the standard picture, after the domination of cosmological constant 
over-densities will grow by a factor of 1.44 and then stop.  Thus
the maximum amplitude achieved by a linear analysis of an rms fluctuation
is
\bea
\sigma_\infty\approx 1.44\times (a_\Lambda/a_{\rm rec})\sigma_{\rm rec}
\approx 5\times 10^{-2}\,\hat{\rho}_{\rm eq}^{1/3}\hat{\zeta}_{\rm eq} \,.
\eea  
Here we have used that 
$\sigma_{\rm rec}\approx 5\times 10^{-5}\hat{\zeta}_{\rm eq}$ on scales 
larger than the Hubble radius at recombination, and that recombination
occurs at a temperature $T_{\rm rec}\approx 3000$ K, where we ignore 
the logarithmic dependence of $T_{\rm rec}$ on $\Mp$ and $\eta$.
The formation of structure still requires that a linear analysis gives 
$\sigma_\infty\geq 1.69$ before the growth in over-densities is halted
by the domination of cosmological constant.  This gives the constraint,
\bea
\hat{\rho}_{\rm rec}\hat{\zeta}_{\rm eq}^{\,\,3} \gtrsim 5\times 10^5 \,.
\label{nsweinberg}
\eea  
Eq.~(\ref{nsweinberg}) constrains $\Mp$ according to
\bea
\left( \frac{1}{6}\hMp^{-\beta}+\frac{5}{6}\hMp^{-1}\right)
\hMp^{-3\alpha} \gtrsim 5\times 10^5 \,.
\label{nsweinberg2}
\eea  
This constraint is much stronger than the constraint it replaces,
Eq.~(\ref{LambdaConstraint}).  

Allowing for baryons to dominate the matter density of the universe 
may affect the other constraints in Fig.~\ref{fig1} in two ways.  First, 
the halo density and background density at virialization, $\rho_{\rm vir}$ 
and $\rho_*$, are now reduced by a factor of $3\times 10^{-6}$ due to the 
difference between $\sigma_{\rm rec}$ and $\sigma_{\rm eq}$.  Second, 
structure formation only occurs on scales greater than the Hubble radius 
at recombination, since sub-horizon perturbations are suppressed.  
Ignoring the $\Mp$ dependence in $T_{\rm rec}$, this implies a 
minimum halo mass set by the horizon mass at recombination, corresponding 
to a scale of roughly $\mu_{\rm min}\sim 10^6$.  It can be shown that no 
value of $\Mp$ satisfies all of the constraints displayed in 
Fig.~\ref{fig1} after these effects have been included.  Dropping the 
constraint that virialization precede recombination does not change this 
result.        

We now turn to the second assumption of Section~\ref{sec:structure}, 
which is that recombination occurs at least an e-fold of expansion 
before virialization.  To investigate what happens when virialization 
occurs before recombination, we adopt the following simplified picture.  
Dark matter over-densities grow when they enter the Hubble radius, and 
we assume that they become non-linear and virialize as they would within 
our universe.  However, growth in the baryon over-densities is hampered 
by their interaction with the photon Hubble flow before recombination.  
Therefore we approximate that baryons do not participate at all in the 
over-densities of the dark matter and are rarefied relative the halo 
density as they follow the Hubble flow.

Within this simplified model, the final baryon fraction within a halo
will be at most about $f_b/18\pi^2$, and will decrease by a factor of 
$e^{-3}\approx0.05$ for each e-fold of expansion between virialization 
and recombination.  However, it turns out that only two of the 
constraints that we consider depend significantly on the baryon 
fraction of the halo.  These are the disk instability constraint of 
Section~\ref{sec:starformation} and the close encounters constraint of 
Section~\ref{sec:closeencounters}.  (The explicit $f_b$ dependence in the 
galactic cooling constraint of Section~\ref{sec:cooling} is canceled by 
an implicit dependence within $\Lambda_c$.)  To explore whether this 
situation opens a new window for allowed values of $\Mp$, it is helpful 
to adopt the following picture.  Instead of simply eliminating the 
recombination timing constraint of Eq.~(\ref{virconstr}), we 
continuously weaken it.  For example, we may demand that recombination 
occur at most $N_{\rm rec}$ e-folds of expansion after virialization and 
then study the above constraints as $N_{\rm rec}$ is increased.  

When we do this, we find that the curves in Fig.~\ref{fig1} 
corresponding to the disk instability constraint, the close encounters 
constraint, and the recombination timing constraint all slide downward 
as $N_{\rm rec}$ is increased.  This shifts the allowed window for $\Mp$
such that larger values of $\Mp$, including the value obtained within 
our universe, become excluded as lower values become allowed.  It 
turns out that the disk instability curve slides downward at a faster
rate than that of the recombination timing curve, so that as the window 
for allowed $\Mp$ moves to smaller $\Mp$ it also grows smaller.  
Ultimately, the window gets pushed against other constraints, such as 
the stellar lifetime constraint or the galactic cooling constraint, and 
disappears.  This happens at about $N_{\rm rec}\approx$ a few.

\section{Analysis of a Structure Formation Constraint}
\label{sec:MpLambda}

In Section~\ref{sec:prior} it is argued that if the probability to 
observe a particular value of $\Mp$ is weighted in part by the 
inflationary expansion factor of universes that contain that value of 
$\Mp$, then it is overwhelmingly preferred that $\Mp$ should be measured 
at one of the boundaries of its anthropic window.  It is clear from the 
discussion of Section~\ref{sec:anthropic} that our value of $\Mp$ is not 
at either of its anthropic boundaries.  Nevertheless, it is worthwhile 
to investigate more quantitatively just how `far' is our value of $\Mp$ 
from its anthropic boundaries.  For simplicity we investigate the 
selection effect from only one anthropic constraint.  Specifically, we 
look at the structure formation requirement that halos virialize before 
the domination of cosmological constant (Section~\ref{sec:ccconstr}).  
Note that this provides the tightest constraint on $\Mp$ 
according to the curves in Fig.~\ref{fig1}.

The arguments of Section~\ref{sec:prior} are appropriate primarily when
the landscape is so densely packed that we can approximate the prior 
distribution $\mI(\Mp)$ to be a continuous function of $\Mp$ within the 
anthropic window.  The probability to observe $\Mp$ to lie within the 
range $d\Mp$ can then be written,
\bea
P(\Mp)d\Mp\propto \mA(\Mp)\mV(\Mp)\mI(\Mp)d\Mp \,,
\eea
where the factors on the right-hand side are defined in 
Section~\ref{sec:prior}.  Of course, we are assuming that universes 
are weighted in part by their inflationary expansion factor $\mV$.  
Since we have no knowledge about the shape of $\mI(\Mp)$, we take 
$\mI(\Mp)\approx$ constant.  As suggested above, we take the anthropic 
factor $\mA$ to be conditioned by only the constraint that halos 
virialize before the domination of cosmological constant prevents this.

\begin{figure}[t!]
\includegraphics[width=8.75cm]{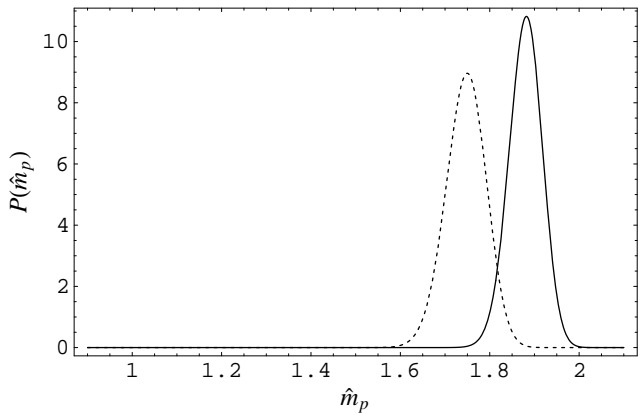}
\caption{The normalized distribution $P(\hMp)$ for $N=60\,\hMp$ 
(solid) and $N=60\ln(\hMp)$ (dashed) for when the landscape distribution
for $\Mp$ depends on the inflationary expansion factor.  Our universe 
corresponds to $\hMp=1$.  In both cases $F$ is defined using $\alpha=3$ 
and $\beta=3/2$.  Although in both cases selection effects appear to 
prefer a specific range for $\Mp$, our value is far outside of this range.}
\label{fig2}
\end{figure}

To proceed, we assume that the likelihood to observe a given value of
$\Mp$ is proportional to the baryon fraction within galaxies of mass
greater than or equal to the mass of the Milky Way, $10^{12}M_\odot$.  
Allowing for smaller galaxies or allowing for observers that do not 
require a galactic environment can only expand the window of allowed 
$\Mp$.  The Press-Schechter function~\cite{virialize} gives the fraction
of matter that collapses into a galaxy of mass greater than or equal to
a given scale.  It is derived in Section~\ref{sec:cc} and given by 
Eq.~(\ref{psfunction}).  We reproduce it here for convenience,
\bea
F= {\rm erfc}\left[ 0.373
\left( \frac{\hat{\rho}_\Lambda}{\hat{\rho}_{\rm eq}\hat{\sigma}_{\rm eq}^3}
\right)^{\!1/3}\right] \,.
\label{massfraction}
\eea   
The prefactor comes in part from evaluating $F$ in the infinite future 
and at the Milky Way mass scale.  The $\Mp$ dependence is given by 
\bea
\left( \frac{\hat{\rho}_\Lambda}{\hat{\rho}_{\rm eq}\hat{\sigma}_{\rm eq}^3}
\right)^{\!1/3}\! = 
\left(\frac{1}{6}\hMp^{-\beta}+\frac{5}{6}\hMp^{-1}\right)^{\!\!-4/3}\!
\hMp^{\alpha} \,.\,
\eea
Note that when $\alpha$ and $\beta$ are positive, $F$ is a decreasing 
function of increasing $\Mp$.  We are interested in the baryonic matter 
within galaxies.  Therefore $\mA$ should contain a factor of the baryon 
fraction $f_b$, given by Eq.~(\ref{baryonf}), along with $F$.  

Finally, we take $\mV(\Mp)\propto e^{3N}$ for $N$ e-folds of inflation.
We require that $N(\Mp)$ be an increasing function of $\Mp$ so that 
$\Mp$ is pushed to larger values, saturating the constraint in 
Eq.~(\ref{LambdaConstraint}).  In addition, we want $\alpha=3$ and 
$\beta=3/2$ so that this constraint on increasing $\Mp$ is as strong as 
possible.  Rather than propose a specific model of inflation, we assume 
that one can contrive a model with the relatively weak dependence 
$N\approx 60\,\hMp$.  Then putting all of our assumptions together gives
the probability distribution,  
\bea
P(\Mp) = {\mathcal N} f_b(\Mp) F(\Mp) e^{3N(\Mp)} \,,
\eea
where ${\mathcal N}$ is a normalization factor.  The normalized 
distribution $P(\Mp)$ is displayed in Fig.~\ref{fig2} for $\alpha=3$ 
and $\beta=3/2$.  Although it is intriguing that in this scenario the 
erfc function overcomes the exponential volume factor, this happens for 
a value of $\Mp$ somewhat larger than the value that we observe.  This
$\mO(1)$ change in $\Mp$ is significant due to the exponential 
sensitivity of $P$ on $\Mp$.  Indeed, the fraction of the distribution 
function $P(\Mp)$ that sits below $\hMp=1$ is completely negligible 
compared to that which sits above (explicitly, this fraction is roughly 
$10^{-53}$). This distribution is so sharply peaked because $N$ is 
relatively large. For example, expanding about the local maximum gives 
$P \sim \exp[-c N \Delta^2]$ where $c\sim\mO(1)$ and $\Delta$ is the 
difference between $\hMp$ and its value at the maximum.  Even if the 
number of e-folds depends very weakly on $\Mp$, for example 
$N\approx 60\ln(\hMp)$, we still find the preference for larger $\Mp$ to 
be overwhelming.  

We now provide a final point of clarification.  A careful reader may 
notice that according to Fig.~\ref{fig2}, the values of $\Mp$ that are 
most likely to be observed lie well outside the anthropically allowed 
windows of Fig.~\ref{fig1}.  This is because the relevant curves in 
Fig.~\ref{fig1} are calculated by assuming that all over-densities have 
initial amplitudes equal to the rms amplitude.  Meanwhile, Fig.~\ref{fig2} 
takes into account that the initial amplitude of an over-density is at
least approximately a Gaussian random variable.  The discrepancy between 
the results in Figs.~\ref{fig1} and~\ref{fig2} reflect that under the 
assumptions of this appendix, the overwhelming majority of galaxies stem 
from over-densities that begin with amplitudes many standard deviations away 
from the norm.  Although these galaxies result from relatively unlikely
initial over-densities, the fact that they arise within enormously 
larger universes more than compensates for this.  This result stems from 
the sharp dependence on $\Mp$ in the inflationary expansion factor.  
If the proper landscape measure does not contain this factor, then the 
distribution for $\Mp$ would be very different than Fig.~\ref{fig2} 
indicates.

\end{document}